\documentclass[10pt]{article}

\usepackage{grffile}
\usepackage{ctable}

\usepackage[small]{caption}
\usepackage{graphicx}

\usepackage{amsmath,amsfonts,amssymb}
\usepackage{amsthm}
\usepackage{color}

\usepackage[letterpaper,textheight=9in,textwidth=7in]{geometry}

\usepackage[skip=10pt,font=footnotesize]{caption}
\graphicspath{{figures/}}

 {\begin{trivlist} \item[]{\bf Proof. }}%
 {\hspace*{\fill}$\rule{.4\baselineskip}{.4\baselineskip}$\end{trivlist}}

 {\begin{trivlist}\item[]\textbf{Acknowledgments.}}{\end{trivlist}}

 {\begin{center}\textbf{Abstract}}{\end{center}}

\makeatletter\@addtoreset{figure}{section}\makeatother

\makeatletter \@addtoreset{equation}{section} \makeatother

\def\XXint#1#2#3{{\setbox0=\hbox{$#1{#2#3}{\int}$ }
\vcenter{\hbox{$#2#3$ }}\kern-.6\wd0}}

\newcommand{\ba}{\begin{align}}
\newcommand{\ea}{\end{align}}

\newcommand{\rmd}{\mathrm{d}}

\newcommand{\rme}{\mathrm{e}}

\renewcommand{\Re}{\mathrm{Re}}

\newcommand{\eps}{{\varepsilon}}

\newcommand{\al}{\kappa}

\begin{document}
\begin{center}
{\fontsize{15}{15}\fontfamily{cmr}\fontseries{b}\selectfont{Signaling gradients in surface dynamics as basis for planarian regeneration}}\\[0.2in]
Arnd Scheel$\,^1$, and Angela Stevens$\,^2$, and Christoph Tenbrock$\,^2$, \\[0.1in]
\textit{\footnotesize 
$\,^1$University of Minnesota, School of Mathematics,   206 Church St. S.E., Minneapolis, MN 55455, USA\\
$\,^2$University of M\"unster (WWU), Applied Mathematics, Einsteinstr. 62,
D-48149 M\"unster, Germany}
\date{\small \today} 
\end{center}

\begin{abstract}
We introduce and analyze a mathematical model for the regeneration of planarian flatworms. This system of differential equations incorporates dynamics of head and tail cells which express positional control genes that in turn translate into localized signals that guide stem cell differentiation. Orientation and positional information is encoded in the dynamics of a long range \textit{wnt}-related signaling gradient. We motivate our model in relation to experimental data and demonstrate how it correctly reproduces cut and graft experiments. In particular, our system improves on previous models by preserving polarity in regeneration, over orders of magnitude in body size during cutting experiments and growth phases. 
Our model relies on tristability in cell density dynamics, between head, trunk, and tail. In addition, key to polarity preservation in regeneration, our system  includes  sensitivity of cell differentiation to gradients of \textit{wnt}-related signals 
relative to the tissue surface. 
This process is particularly relevant in a small tissue layer close
to wounds during their healing, and modeled here in a robust fashion  through dynamic boundary conditions. 
\end{abstract}

\setlength{\parskip}{4pt}
\setlength{\parindent}{0pt}

\section{Introduction}

%
%
%
%
%

Planarians are nonparasitic flatworms commonly found in freshwater streams and ponds \cite{reddien2004fundamentals,bookrink}
with a body size in the \textit{mm}-scale. They possess the ability to regenerate after rather severe injuries to their body. In extreme cases, the entire organism can regenerate perfectly from a fragment just $0.5$\% of the original size. 
The potential of the relevant processes 
has helped direct a tremendous amount of attention toward studying mechanisms during regeneration in these organisms. 

The purpose of this paper is to present a minimal model, informed by experimental data, that reproduces this fascinating regenerative behavior. 
To the best of our knowledge, the mathematical model we introduce here, 
is to date the only one, based on reacting and diffusing species, that is able to correctly recover most of the typical cutting and grafting experiments. In particular, our model shows preservation of polarity during regeneration small tissue fragments.  This reflects 
experimental findings where small tissue parts cut from a planarian 
regenerate to a fully functioning and intact organism with  
head and tailpositioned such, that the 
original orientation of the tissue fragment is respected.
On the other hand, when tissue parts are cut from a donor and implanted into a host, the newly created planarian integrates the positional information of the
tissue fragment from the donor with the new positional information it obtains from the host. \\
The  mechanism proposed here resolves a conundrum in modeling efforts. In fact, many models of spontaneous formation of finite-size 
structure in unstructured tissue allude to a Turing type mechanism 
to select a finite wavelength.  
These types of models have often been discussed in the context of
regeneration phenomena in hydra, which  exhibit similar robustness features 
as planarian regeneration that we focus on here. 
Regeneration of hydra has been one of the central motivations for many studies 
of activator-inhibitor type systems \cite{mh407}. 
Such Turing mechanisms however do not scale across several orders of magnitude as seen in the patterning of planarians, nor do they incorporate robust selection of polarity. In the remainder of this introduction, we briefly describe the species and kinetics in our model, the role of dynamic boundary conditions, and the key experiments that we aim to mimic in our modeling efforts. 

\paragraph{Cell types and signaling agents.}

We focus on the ante-posterior axis, describing dynamics in a one-dimensional domain $x\in [-L,L]$, populated by different cell types and signaling agents. 
Head cells $h$ and tail cells $d$ are generated from stem cells $s$ through differentiation. Head and tail cells generate signals $u_h$ and $u_d$, respectively. In addition, we model production and diffusion of a long-range 
\textit{wnt}-related signal $w$, in the following often referred to as  $\textit{wnt}$-signal in short, that encodes orientational information through its gradient. We model random motion of cells and diffusion of signal molecules 
$U(t,x)=(s,h,d,u_h,u_d,w)(t,x)$ through diffusion coefficients $D_j$, and postulate kinetics  
\begin{itemize}
 \item $F_s(s,u_h,u_d)$ for  proliferation and differentiation of stem cells;
 \item $F_h(s,h,u_h),F_d(s,d,u_d)$ for differentiation of stem cells into head and tail cells and death of those;
 \item $F_{u_h}(h,d,u_h),F_{u_d}(h,d,u_d)$ for regulation of the signals $u_h$ and $u_d$ by head and tail cells;
 \item $F_w(h,d,w)$ for degradation and production of the \textit{wnt}-related signal by head and tail cells, respectively. 
\end{itemize}
The kinetics of $F_s$ regulate a near-constant supply of stem-cells. 
The dynamics for the signals $u_h$ and $u_d$, which are produced by head and 
tail cells but degraded by reactions with tail and head cells, respectively, 
encode tristability between a head-only, a tail-only, and a zero state. The 
long-range signal $w$ is produced by tail cells and degraded by 
reactions with head cells. As a result, a healthy planarian consists of a 
high concentration of $h$ near $x=-L$, a high concentration of $d$ near 
$x=L$, and a near-constant gradient of $w$ in $x\in(-L,L)$, see  
Figure \ref{f:schematics}.

\paragraph{\textit{wnt}-related regulation of cell types.} We emphasize here that differentiation rates do \emph{not} depend on $w$ in our model. In other words, we do not incorporate mechanisms that convert levels of 
the \textit{wnt}-related signal (or other long-range signals) into positional information that in turn  directs the differentiation process. Such mechanisms, often referred to as the French-flag model \cite{WOLPERT19691,Wolpert94}, are widely assumed to play a key role in early development and likely also have significant influence on the regeneration of planarians \cite{bookrink}. Here, we omit 
this dependence for several reasons. First, we believe that such positional 
information is inherently inapt at explaining the preservation of polarity, 
since the absolute levels of the \textit{wnt}-signal play apparently little 
role in the regeneration from small tissue fragments. This is exemplified by 
the robustness of regeneration and polarity in regards to the location of the 
cut out tissue fragment, near head, near tail, or from central parts of the 
body. Second, such information is not necessary in the early stages of the 
regeneration, but seems to be relevant only later, when also the size of functional regions such as head or tail are regulated. 
Lastly, related to this latter point, there is little concrete information 
so far on the nature of biological mechanisms that would accomplish this 
translation of, for instance,  \textit{wnt}-signal levels into an information 
for differentiation of stem cells in a robust fashion. 
Therefore we see value in not simplifying the modeling task by the addition of such somewhat poorly substantiated regulatory mechanisms.

Rather than using positional information based on the absolute value
of the \textit{wnt}-signal, our mathematical model is crucially based on the 
detection of gradients of \textit{wnt}, in particular the orientation of the 
gradient relative to the
respective body edge (encoded in the outer normal of the boundary). We see 
this orientation of the gradient as an in some sense necessary, minimal information, to guide regeneration while at the same time preserving polarity. This 
detection of gradient orientation is encoded in boundary dynamics that we shall explain next. 

\paragraph{Boundary dynamics.} 
Cutting experiments eliminate head and/or tail cells, therefore ultimately 
influence the \emph{wnt}-pathway and destroy the associated signaling gradient. We therefore include boundary dynamics into our model which represent wound 
healing processes at wound edges that in particular lead to regeneration of 
head, tail, and ultimately reestablish the signaling gradient of $w$. 
This is due to fast differentiation of stem cells close to the 
wound edges. Differentiation into head versus tail cells is guided by 
information on the normal derivative of the residual signal $w$ in the body fragment. We model these boundary dynamics through dynamic boundary conditions (sometimes referred to as Wentzell boundary conditions in the literature). Therefore, we introduce values $U_\pm(t)$ of all cell and signal concentrations at the left and right boundary compartment, respectively, that specifically act as Dirichlet conditions for the chemical species $U(t,x)$. 
We then introduce differential equations for $U_\pm(t)$ based on a diffusive flux $\partial_\nu U$ to ensure mass balance, the \emph{same} kinetics $F_j$, now evaluated at the boundary, and, crucially, one additional term that accounts for strong differentiation of stem cells into 
head and tail cells, namely $s\cdot\Psi_{h/d}$, during wound healing. We postulate that,  as a key ingredient to robust regeneration and preservation of polarity, the production rate $\Psi_{h/d}=\Psi_{h/d}(h,d,\partial_\nu w)$ depends sensitively on the normal derivative of the long-range signal $w$, measuring roughly the sign of $\partial_\nu w$. We model this switching behavior through a rate function
\[
 \Psi_h(h,d,\partial_\nu w)=\tau(1-h)(1-d) \chi_{>\theta}^\eps(\partial_\nu w),\qquad
 \Psi_d(h,d,\partial_\nu w)=\tau(1-h)(1-d) \chi_{<-\theta}^\eps(\partial_\nu w),
\]
where $\tau\gg 1$ is the rate and $\chi_j^\eps$ are smooth versions of the characteristic function such as
\begin{equation}\label{e:char}
 \chi_{>\theta}^\eps(\xi)=\frac{1}{2}\left(\tanh((\xi-\eps\theta)/\eps)+1)\right),   \qquad
 \chi_{<-\theta}^\eps(\xi)=\frac{1}{2}\left(\tanh(-(\xi+\eps\theta)/\eps)+1)\right);
\end{equation}
see Figure \ref{f:chi} for an illustration. The steepness $\eps^{-1}$ of the smoothed characteristic function can be interpreted as a sensitivity of the production in regards to small gradients. 
One can envision many scenarios that enable stem cell differentiation to be guided by gradients of a chemical signal, for instance through comparing signal strength spatially or temporally. This latter process can be enhanced by directed motion of stem cells or progenitors into the respective directions. Indeed, migratory stages of progenitors from their place of birth to their site of terminal 
differentiation are discussed in \cite{bookrink}.

From a mathematical perspective, it is interesting to note that dynamic boundary conditions of this type cannot be readily replaced by, say, Robin boundary conditions, by letting for instance rates of boundary dynamics tend to infinity. This mathematical curiosity bears consequences on modeling assumptions, implying for instance the presence of a distinguished boundary/body region. Such a region has recently also been 
discussed in the experimental literature \cite{bookrink}, where
the concept of poles, separating head regions from trunk, is attributed 
a key role in regeneration. On the mathematical side, we make this curious role of dynamic boundary conditions precise  in a simple reduced model toward the end of this paper.  

\paragraph{Regeneration from cutting, grafting, and growth --- simulations.}
The model ingredients presented thus far can, to a large extent, be motivated through experimental data. We validate the modeling effort by displaying results from numerical simulations that mimic basic experiments on regeneration, such as cutting, grafting, and growth; see Section \ref{s:3}. 


\paragraph{Regeneration --- analysis.} We provide some analytical 
understanding of the key ingredients of our mathematical model by deriving a 
reduced system, consisting of an order parameter $c$ that lumps concentrations of head and tail cells into a scalar quantity, coupled to the long-range \textit{wnt}-related signal $w$. In this reduction, we clearly illustrate how the nonlinear boundary fluxes restoring head and tail cell concentrations, 
dependent on the information of normal derivatives of the gradient, initiate and organize the regeneration process. We outline a (in)stability analysis that points to the origin of regeneration and we pinpoint failure of this process when dynamic boundary conditions are relaxed to, say, nonlinear mixed (or Robin) 
boundary conditions. Lastly, we distill the key feature, restoration of a 
signal gradient through boundary, respectively body edge, sensing, into a dramatically oversimplified scalar model for the \textit{wnt}-related signal.  In this scalar  model, we then clearly outline the limits of regeneration and point to curious oscillations, caused by coupling of scalar, non-oscillatory dynamics in the boundary to the diffusive signal field in the bulk and the resulting delayed feedback mechanism. 

\paragraph{Outline.} We review biological experiments that motivate our model  and compare with other systems in the literature in Section \ref{s:2}. We then 
introduce our mathematical model and describe numerical simulations that mimic planarian regeneration in various scenarios in Section \ref{s:3}. Section \ref{s:4} contains analytical results that reduce dynamics to a two-species and ultimately a scalar reaction-diffusion system, thus exhibiting the key dynamic ingredients of our original model. 
We also discuss the necessity of the somewhat non-standard boundary conditions and the limitations of the tristability mechanism. 
We conclude with a brief summary and discussion.

\paragraph{Acknowledgments.} 
CT was supported by a fellowship of the Graduate School of the 
Cells-in-Motion Cluster of Excellence EXC 1003--CiM, 
University of M\"unster (WWU). 
ASc was partially supported through NSF grant  DMS--1311740 and DMS--1907391, a
Research Award 
from the Alexander-von-Humboldt Foundation,  and a WWU Fellowship.
ASc and ASt were supported by the DFG 
(German Research Foundation) under Germany’s Excellence Strategy 
EXC 2044-390685587, Mathematics M\"unster: Dynamics -  Geometry – Structure.
CT and ASt gratefully acknowledge numerous valuable discussions about 
the biology
of planarians with Kerstin Bartscherer. 

\section{Planarian regeneration --- experiments and models}\label{s:2}

We give a brief overview of planarians in Section \ref{s:21},  
describe experiments with a short comparison to our findings from
the mathematical model 
 in Section \ref{s:exp}, give genetic information in Section \ref{s:gen}, and summarize the existing mathematical literature in Section \ref{s:mat}. 

\subsection{Basics of planarian regeneration}\label{s:21}

In planarian pluripotent adult stem cells (neoblasts), are the source of all 
cell types. Neoblasts divide, differentiate, and differentiated 
cells die after some time. There is a continuous turnover of cells, still the 
flatworm maintains cell type proportions. 
Understanding the mechanisms responsible for this dynamic steady state is 
one important objective in research on planarians. 



Planarians exhibit a bilateral symmetry, three distinct body axes, a 
well-differentiated nervous system including a brain, a gastrovascular tract, 
a body-wall 
musculature, and they consist of three tissue layers \cite{lobo2012modeling}. 
The best experimental data is available for the species "Schmidtea mediterranea", which is $1\,\textit{mm}$ to $20\,\textit{mm}$ long \cite{reddien2004fundamentals} and 
consists of $100.000$ to more than $2.000.000$ cells. At least $20$ to $30$ different 
types of differentiated cells are observed. The broadly distributed 
pluripotent population of neoblasts constitutes 
possibly up to 
$30\,\%$ of the total cell population \cite{Baguna3} and are the only 
proliferating cells in planarians  \cite{tasaki2011erk}. 

Regeneration in our context is the ability of an organism to replace lost or damaged tissue. 
The regenerative abilities of humans are limited to special tissues
only. 
Examples for regeneration after injury  in animals 
include regeneration of deer antlers, fins of fish, tails of geckos, or complete limbs in some crabs or salamanders. Planarians and hydra are among the few species that seem to possess a nearly unlimited regenerative ability. They recover from practically every injury, and regenerate when aging.  A small fragment 
cut out  of a planarian, as small as about 0.5\,\%,
can regenerate a full animal, including in particular an intact brain \cite{reddien2004fundamentals}. Some asexual planarian species even reproduce by tearing themselves apart and subsequently develop into two intact worms \cite{lobo2012modeling}. 
Clearly, the study of this behavior holds tremendous appeal when considering the potential of  a  better understanding of regeneration of  tissues in the human body, for instance parts of the heart muscle after an infarct. 

\subsection{Experiments}\label{s:exp}

We give a short overview of classical experiments and refer to 
\cite{baguna1981quantitative,baguna1999morphology,baguna1976mitosis,bardeen1904inhibitive,bowen1974fine,de1984chromosomal,salo1985cell,morgan1904control,randolph1892regeneration,randolph1897observations,santos1929studies,smales1978epidermis,newmark2002not,morgan1905polarity,morgan1898experimental,salo1984regeneration}
for more details. We also briefly point to and summarize our findings in relation to these experiments. 

\subsubsection{Cutting experiments}
\begin{figure}[!ht]
   \centering
    \includegraphics[width=0.8\textwidth]{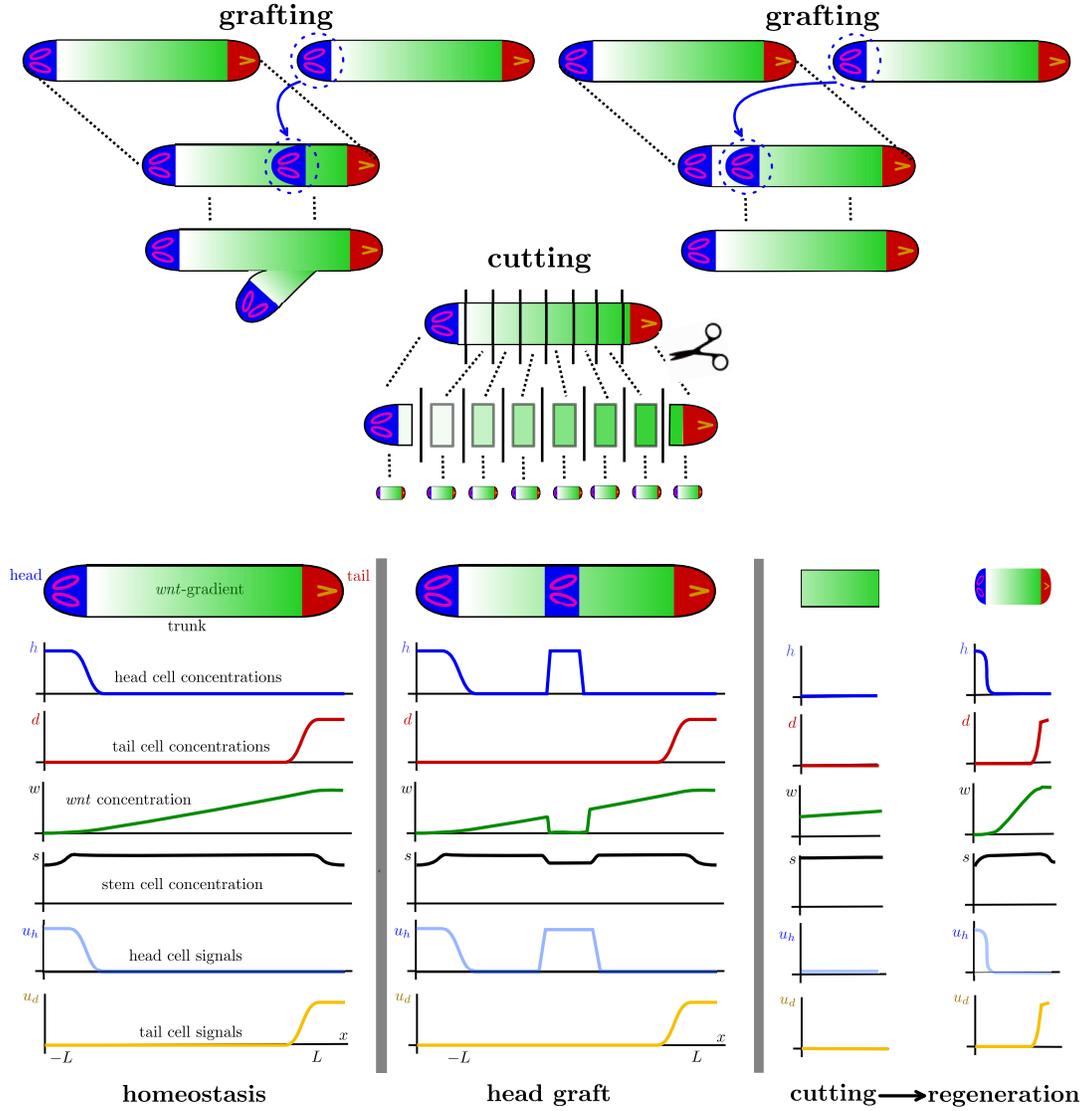}
   \caption{Typical experiments and their representations
in terms of our mathematical model.  Top left: Schematic illustration of 
head grafting and regeneration of a two-headed animal. 
Top right: Head grafting and regeneration into a normal planarian with 
one head and one tail.
Middle row: Cutting experiment and regeneration of 8 planarians.  Bottom row: Schematic illustration of experiments on planarians and 
associated spatial distribution of concentrations $h,d,w,s,u_h,u_d$. 
Homeostasis (left) has head cells concentrated on the left, tail cells concentrated on the right, a \textit{wnt}-gradient directed towards
   \emph{wnt}-production in the tail, roughly constant stem cell populations 
with slightly decreased populations where differentiation into head and tail 
cells occurs, and head and tail signals are closely mimicking the distribution of 
head and tail cells. In grafting experiments (center), the head region of a 
donor   
planarian is grafted into a host planarian, retaining roughly the distribution of 
cell concentrations and signaling molecules from its original location. In 
cutting experiments (right), a thin fragment (here from the trunk region) 
retains a small \textit{wnt}-gradient but no head or tail cells. Regeneration 
in this context refers to reestablishing head and tail cell populations 
while preserving polarity.}\label{f:schematics}
\end{figure}

Cutting tissue off a planarian results in regeneration of both parts
into an intact organism, more or less independent of position, size, or direction of the cut, with few exceptions, discussed below. Here we focus on regeneration of the anterior-posterior (AP, or head-to-tail) axis, and think mostly of dissections transverse to the AP axis into strips whose width is a fraction of the 
original length of the planarian body; see Figure \ref{f:schematics}. 
Each of these tissue strips will regenerate into a complete animal after several weeks. In tissue pieces cut close to head or tail, regeneration of a new head might occur in as little as three to four weeks. Complete regeneration, 
including restoration of the right proportions, usually requires two to three 
months, depending on the shape of the respective tissue fragment  \cite{morgan1898experimental}; see Figure \ref{f:schematics} for schematics of experiments and representation in our model. 
Polarity in such fragments is preserved, that is, a head  will regenerate at 
the edge that has been closer to the head before, and the tail will regenerate 
at the opposite end. In particular, neighboring cells in a planarian will 
regenerate as either head or as tail after they have been separated by a cut!
Among the few  exceptions are very long or very short tissue fragments, which sometimes regenerate a second head instead of a tail. Moreover, tissue parts that are too small 
or too thin are not able to regenerate at all.

Our mathematical model reproduces all of these experiments robustly, for a 
wide range of parameter values, with a few exceptions that we note below; 
see Figure \ref{f:cutting} for simulations of our full system and 
Figures \ref{f:red}, \ref{f:red2}, and \ref{f:scal1}, \ref{f:scal2} for 
simulations of a reduced model and for a scalar equation. 
In our mathematical model head and tail cells 
are treated equivalently with equal weights to cells and corresponding reaction rates. As a consequence, simulations are not able to produce a bias between two-headed 
and two-tailed animals. Specific outcomes depend only on small fluctuations 
in initial conditions when recovery is not robust, that is for very small or 
or for large fragments. Matching experiments, we found recovery somewhat more 
robust for tissue fragments from the center of the body when compared to 
equally sized fragments cut from regions near head or near tail. In the 
analytically best 
understood case of the scalar model, we observed that
when changing parameters 
toward a regime where regeneration is impossible, oscillations occure in the recovery process before, upon further change of parameter values, the \textit{wnt}-gradient fails to recover at all. These are the exceptions from the otherwise 
matching behavior of our mathematical model in comparison to experimental findings. 


\paragraph{Grafting experiments.}

In grafting experiments a tissue fragment of an intact donor planarian is cut 
out and transplanted into  another healthy host planarian, whose regeneration 
is then studied. The donor usually regenerates as described before, whereas 
the host may develop new phenotypes; see again Figure \ref{f:schematics}. Most interesting are experiments where parts of the head of the donor are 
transplanted into different positions within the host. Depending on size and position of the 
donor head tissue within the host, the host will either regenerate into a 
completely normal animal, with the grafted head 
vanishing, or it will generate a second axis at the position of the transplant such that the AP axis splits into two branches 
of equal length exhibiting two heads and one tail, or it will generate a 
two-headed planarian. Data suggests that outgrowth is more likely 
if there is a larger
distance between the head transplant and the head of the host. Similarly, 
outgrowth seems to become more likely for larger head transplants and for tissue taken closer to the head of the donor.

Transplanting a donor tail into the upper part of the host might reorganize the host's middle region, sometimes regenerating a second pharynx in opposite 
direction, possibly also leading to outgrowth. Transplanting a complete 
lower donor fragment below an upper fragment of a bisected host, such that 
just an intermediate strip for a complete animal is missing, will always 
regenerate the intermediate strip. 

We mimic such experiments in simulations of our mathematical model; 
see Figure \ref{f:grafting} and \ref{f:graftingrand} for the full
system and Figure \ref{f:red} for the reduced model and confirm to some extent the observed dichotomy between the emergence of a new head and the vanishing of the grafted head during regeneration. As a caveat, the insight into these phenomena from our model is limited by the fact that we do not model regulation of organ growth through chemical signals.


\paragraph{Growth and Shrinking.}

Adult  planarians of the species ``Schmidtea mediterranea'' are roughly $20\,\textit{mm}$ long if well fed, and can shrink to about $1\,\textit{mm}$  if starved, and regrow when food supply is restored, keeping relative proportions
and ratios of cell populations intact \cite{Baguna3}. During regeneration after cutting experiments, this ability is crucial: if, for instance, a tissue fragment is cut from close to head or tail, it will initially lack a pharynx, and therefore draw resources from the fragment shrinking to as little as a tenth of the original fragment size. 

We show experiments that show robust behavior over a wide range of body sizes and  growth rates; see Figure \ref{f:growth}.

\paragraph{Hydra.}

The regeneration of hydra, a freshwater polyp, shows similarities to that of planarians, with comparable outcomes for most of the experiments mentioned above; see \cite{webster1966studies,macwilliams1983hydra,shostak1972inhibitory,shimizu1993minimum,bode2003head,wilby1970experimental} and references therein. 
Different from planarians, hydra stem cells are distributed exclusively inside the body region, below the epithelium, such that a fragment consisting only of head or foot cells will not regenerate; see \cite{achermann1985genetic} for one of the very few quantitative studies on induction of additional foot or tail axes.

An interesting \textit{dissociation experiment} for hydra has not been described for planarians so far. 
When a tissue fragment of hydra is pressed through a net and the resulting small 
fragments are reorganized randomly, a bulb of hydra tissue without any sense of polarity and positional information arises. This tissue fragment will subsequently regenerate head and foot structures.  However, depending on the number of 
involved cells, it might generate several heads and body axes that will separate only later \cite{noda1971hydra}; compare Figure \ref{f:graftingrand} for a related set-up
in our model.
%

\subsection{Genetic information.}\label{s:gen}

%
In order to identify gene expression and their spatial and temporal 
dynamics in planarians, the amount of RNA  produced from 
specific genes during protein synthesis is measured, for instance through 
in situ hybridization or northern blot \cite{nusslein1980mutations}. Therefore  
genes can be identified, which are mainly expressed close to the head 
or tail, or after wounding and feeding. To artificially influence 
gene expression by RNA interference, targeted mRNA molecules are neutralized, 
thus synthesis of messenger molecules stops at an earlier point in time. 
With this 
manipulation it can be tested, e.g. which
of the genes being expressed within the head region are actually necessary to regenerate a head.

So far, in  both hydra and planarians,  it does not seem possible to track the corresponding messenger molecules directly. Corresponding antibodies are not yet available. The experimental methods described above, do allow however, 
to some extent, for an analysis of the production dynamics and functioning of 
these signals. The identification of their final distribution in the 
planarian body
seems out of reach at the moment, but would be important to know for 
further mathematical modeling efforts. 
We refer to  \cite{REG2:REG256}, \cite{bookrink} for a review on biological 
results pertinent to our mathematical model presented here. 

\paragraph{Gene expression sites.}
The three main body axes in planarians are organized by different signaling systems. Knock-out experiments  suggest that these systems act quite independently \cite{reddien2011constitutive}. We therefore focus on genes which organize the anterior-posterior (AP) axis. Information about the other axes, and more details on the AP axis can be found in \cite{adell2010gradients,reddien2011constitutive,petersen2008smed,gurley2010expression, almuedo2012wnt,gagliardi2014inhibitors,reuter2015beta}.
\\
Genes which are expressed close to the head are \textit{notum, sFRP-1, ndl-4, 
prep}, and \textit{sFRP-2}, among others. While being expressed mainly in the upper half of planarians, some of these genes are also expressed at the lower tip of the pharynx region in the center of the body. Genes which are expressed closer to the tail include \textit{wnt1, fz4, wnt11-2, wnt11-1, wnt11-5 (or 
wntP-2)}, 
and \textit{ Plox4-Dj}. 
Among the head related genes, \textit{notum} and \textit{sFRP-1} are expressed more locally, while expression of \textit{sFRP-2} extends from the planarian 
head tip to its center. Similarly, among the tail related genes, \textit{wnt1} is expressed very locally, while \textit{wnt11-5} is expressed from the tail 
tip to the center region in a graded fashion. Local expression here refers to 
only a few cells at the very tips of the planarian that are responsible for 
the production of associated molecules, most likely subepidermal muscle cells \cite{witchley2013muscle}.

The time dynamics of gene expression 
after wounding are described in \cite{almuedo2012wnt,petersen2009wound,wenemoser2012molecular,petersen2011polarized}. When a tissue strip of a planarian 
is cut from the center of the body, 
lacking head and tail, one can distinguish between genes that are activated early,  during the first $0-18$ hours after amputation, or later. 
Further one can identify genes that are expressed asymmetrically, either at 
head or at tail wounds but not at both. Genes, that are expressed very early 
after wounding include \textit{wnt1, wnt11-5, notum, sFRP-1}, but only \textit{notum} is expressed asymmetrically, that is, at wounds that face missing head 
structures. Genes such as \textit{wnt1} are expressed first at all wound sites 
and their expression is normalized to the behavior we described before, only 
later. Among $128$ wound induced genes only \textit{notum} shows the 
polarized expression \cite{wurtzel2015generic}, mentioned already, and only the downstream factor of \textit{wnt}-signaling, $\beta$\textit{-catenin}, seems to influence this asymmetric expression \cite{scimone2014forkhead}. 

\paragraph{Knock-out experiments.} 
While it is highly desirable to acquire information about the action of 
genes and especially their corresponding messenger molecules,
a clear picture seems elusive at this point,
due to the shear number of genes involved in regeneration 
processes; see  
 \cite{reddien2011constitutive} and \cite{REG2:REG256} for recent reviews. 

Inhibition of most genes that are expressed early after wounding  with expression patterns related to head or tail identities causes regeneration failures of the respective body region at their sites of expression. For instance, inhibition of \textit{notum} and head amputation prevents head regeneration, while inhibition of \textit{wnt1} and tail  amputation prevents  tail regeneration. In these cases the planarian will regenerate with two tails or two heads, respectively. 
Considering \textit{wnt1} and \textit{notum} as one of the first expressed 
genes after wounding it is interesting to note that they act in an 
antagonizing manner \cite{kakugawa2015notum}. 

Most important for establishing and maintaining the AP axis/polarity 
appears to be the canonical \textit{wnt}-pathway. This can be demonstrated by influencing  the level of $\beta$\textit{-catenin} inside the cell. 
High levels of \textit{wnt} signaling correlate with high levels of $\beta$-catenin within the cytoplasm. 
The inhibition of $\beta$\textit{-catenin} itself leads to different phenotypes
 \cite{adell2010gradients}. If low doses of inhibitor \textit{dsRNA} are injected 
into a planarian and the tail is removed, the wound closes but no tail 
regenerates. On the other hand, higher doses will lead to regeneration of a 
second head at the tail wound with a second opposing pharynx in the middle
 of the planarian body. If the doses are further increased, the pharynx (both) become disorganized and ectopic eyes appear. A complete inhibition of $\beta$\textit{-catenin} leads to a radially shaped planarian which has head-related structures (nerve cells, eyes, etc.) at every side, even without tail amputation. In this context, we remark that more recently,  an organizing $\beta$\textit{-catenin} concentration gradient, with maximum at the head and minimum at the tail  has been confirmed experimentally \cite{sureda2016localization,stuckemann2017antagonistic}.

\paragraph{Gene expression in hydra.}
The hydra homologue of the above mentioned genes \textit{wnt, dishevelled, gsk3, tcf} and $\beta$\textit{-catenin} seem to act in a comparable way. Studies on expression pattern and knock-out experiments can be found  in \cite{plickert2006wnt,hobmayer2000wnt,philipp2009wnt,lengfeld2009multiple,gee2010beta}. The similarities appear strong enough to suggest that models developed here for planarians would carry implications also for hydra. More generally,  the \textit{wnt}-pathway appears to be more widely conserved during evolution in many species, beyond hydra and planaria.

\subsection{Mathematical models in the literature}\label{s:mat}
Efforts toward mathematical modeling the emergence of pattern or structure in 
seemingly unpatterned developmental stages of organisms 
go back  to Alan Turing's seminal work \cite{Turing}, where he  studied the 
possibility of pattern formation in systems of reaction-diffusion equations, 
based on disparate diffusion length scales. He included in particular the case 
of vanishing diffusion and elaborated on the selection of 
long-, \mbox{finite-,} and short-wavelength patterns, as well as oscillatory, 
traveling-wave patterns, depending on reaction constants. The key observation 
was that it is possible that the spatially extended system can be unstable, 
even when simple kinetics do not exhibit instability, hence the terminology of 
diffusion-driven instabilities and pattern formation. He also modeled the 
development of tentacles in hydra \cite{Turing} but could not finish a 
follow-up manuscript on mathematical modeling of developmental processes 
because of his untimely death. The system of two reaction-diffusion equations 
from Turing's paper has been tremendously influential in the literature and 
was used and built upon by others to model, for instance regeneration in hydra, 
in particular in  \cite{gierer1972theory} where the notion of 
activator-inhibitor system was introduced. For other models based
on such kind of dynamics see \cite{murrayI}, \cite{murrayII} and the references therein. 
Turing's mechanism has also been incorporated into descriptions and models of 
regeneration after cutting experiments in hydra and planarians and a certain 
class of grafting experiments. However, the main feature of Turing's 
mechanism, the selection of a preferred finite wavelength, like for the
tentacles of hydra, there turns into a drawback since patterns in such models 
strongly depend on the domain size, exhibiting  more concentration maxima and 
minima on larger domains, while the pattern in the regeneration 
of the planarian body axis 
develops and is robust over several orders of magnitudes of variation in body size. Attempts in the literature to address this conundrum are typically built on additional species that evolve dynamically and affect reaction constants in Turing's mechanism in ways that change Turing's selected wave length. In well designed contexts, one can thus achieve scaling invariance of patterns over large changes in domain size \cite{Othmer4180,Umulis4830,umulis09,friedrich}, thus providing mechanisms and models for experimental evidence of scale-independent patterning as for instance in zebra fish \cite{Almuedo-Castillo2018}.

With a focus back to planarians, regeneration of a pattern would not necessitate the formation of multi-modal structures but rather necessitate establishing a monotone signaling gradient. Generation of such monotone structures has in turn been studied mathematically quite extensively in other modeling contexts, for instance in phase separation \cite{MR1772733} and in cell polarization \cite{mori2008wave}. In the reaction-diffusion context, the simplest formulations lead to 2-species mass-conserving reaction-diffusion systems, which can exhibit pattern formation, albeit with a wavelength proportional to the domain. Patterns of smaller wavelength are unstable against coarsening although coarsening can be slow or even arrested for small or vanishing diffusivities. 

Motivated by the receptor-ligand binding model \cite{sherratt1995receptor}, a more complex
model for regeneration in hydra was developed  in  \cite{marciniak2003receptor}, again based on 
diffusion driven instabilities with features of activator-inhibitor models. Later, introduction of hysteresis into such models allowed the authors to correctly recover grafting experiments in numerical simulations of  this system of six coupled ordinary and partial differential equations for most relevant scenarios  \cite{marciniak2006receptor}. Intuitively, mechanisms based on hysteresis or multistability are well suited to describe the outcome of grafting experiments as higher values of suitable variables, which 
correspond to head identities, are stabilized, independent of position or 
domain size. Additionally, effects such as the absorption of very small transplants
might be explainable, too. In fact, our model contains some features of the associated multistability. On the other hand these models typically do not generate relevant patterns robustly in situations corresponding to cutting and dissociation experiments. Systems of ordinary differential equations (vanishing diffusivities) coupled
to partial differential equations have been studied extensively, yet more recently in regard to their pattern-forming capabilities  \cite{marciniak2017instability,marciniak2013dynamical,harting2014spike,marciniak2016diffusion,li2017bifurcation,marciniak2015pattern}, finding for instance stable patterns and unbounded solutions developing spikes. 
In a different direction, the role of hysteresis in diffusion-driven instabilities and de novo
formation of stable patterns was investigated  in \cite{harting2015stable,marciniak2017fitzhugh-nagumo} 

Lastly, we point out that our analysis connects with recent efforts to model and understand the role of distinguished surface reactions and bulk-to-surface coupling in morphogenesis \cite{elliott,CUSSEDDU2018,MR2203643,MR2993944,MR3246156}. Envisioning for instance two species reacting and diffusing with equal diffusion constant on a surface, but one of the species diffusing rapidly into, through, and back out of the bulk, one immediately finds the disparate effective diffusivities necessary for pattern formation in Turing's mechanism, thus providing a biologically realistic mechanism for robust pattern formation. 

We refer to \cite{tenbrock} for a more in-depth discussion of recent models 
based on diffusion-driven pattern formation, and in particular also to the  
modeling and simulation of 
some variants of the mathematical model discussed here. 

As we shall see below, our model bears little resemblance with the modeling 
efforts described in this section. It is arguably simpler, at least in the 
reduced forms of Section \ref{s:4}, than most models based on
diffusion-driven instabilities, and possibly more versatile in the experimental phenomenology that can be robustly reproduced.

\section{Cell types, signals, and boundary dynamics}\label{s:3}

We introduce cell and signal densities motivated by the discussion in 
Section \ref{s:2}, together with the associated production rates in Section \ref{s:3.1}. We then discuss dynamic boundary conditions and the additional gradient-sensing effect in Section \ref{s:3.2}. Section \ref{s:3.3} contains numerical results on cutting, grafting, and growth, illustrating the features of our 
mathematical model. 

Much of the mathematical formulation, including cell types and signals, is based on  the thesis of one of the authors \cite{tenbrock}, but we deviate in our treatment of boundary conditions and gradient sensing.

\subsection{Cell types, signals, and rates in the bulk}\label{s:3.1}
We restrict ourselves to processes regulating the 
anterior-posterior (AP) axis and therefore consider a simple one-dimensional spatial domain $x\in[-L,L]$. 

\paragraph{Cell types.} We model three {cell types} depending on time $t$ 
and position $x$ within the planarian, \\[4mm]
\hspace*{1.5cm}\begin{tabular}{p{1.5cm}p{7cm}}
\toprule
 $s(t,x)$ & density of stem cells\\ 
 \midrule
 $h(t,x)$ & density of head-specific cells\\
 \midrule
 $d(t,x)$ & density of tail-specific cells\\
 \bottomrule
\end{tabular}
\paragraph{Signals.} 
Next, we distinguish between signals, whose actions are strongly associated with one of the cell types and are typically very localized, and signals with long-range effects that are influenced by more than one cell type. Short-range signals in our model, acting locally, are\\[4mm]
%
%
%
%
\hspace*{1.5cm}\begin{tabular}{p{1.5cm}p{7.cm}}
\toprule
 $u_h(t,x)$ & density of signals related to head-specific cells \\ 
  \midrule
$u_d(t,x)$ & density of signals related to tail-specific cells\\
 \bottomrule
\end{tabular}\\[4mm]

In addition, we consider a longer range \textit{wnt}-related signal that 
establishes a full body gradient. It is produced by tail cells and  
degraded
by reactions with 
head cells,\\[4mm]
\hspace*{1.5cm}\begin{tabular}{p{1.5cm}p{7cm}}
\toprule
 $w(t,x)$ & density of lumped wnt-signaling\\
 \bottomrule
\end{tabular}\\[4mm]
The idea here is that body fragments retain a gradient in the concentration of this signal $w$ which can then provide clues for polarization in regeneration. 
In fact, much of the current understanding of planarian regeneration rests on the idea of such 
a full body gradient of \textit{wnt}-signaling \cite{adell2010gradients}, although details are not completely understood. Measurements of $\beta$\textit{-catenin}, a cell-internal downstream 
factor of \textit{wnt}-signaling \cite{gagliardi2014inhibitors} seem to support this idea of a long-range gradient in concentration, decreasing from tail to head \cite{stuckemann2017antagonistic,sureda2016localization}. Expression of 
\textit{wnt1} is clearly associated with tail identities \cite{petersen2009wound} and leads to accumulation of $\beta$\textit{-catenin} inside the 
cells \cite{cramer2004untersuchungen}. On the other hand \textit{notum}, a \textit{wnt} antagonist \cite{kakugawa2015notum} expressed locally at the tip of 
the head \cite{reddien2011constitutive}, is the only gene that shows polarized
expression directly after wounding \cite{wurtzel2015generic} and is 
required for head regeneration \cite{scimone2014forkhead}. Its expression
is exclusively affected by $\beta$\textit{-catenin} signaling \cite{scimone2014forkhead,petersen2011polarized}.
It therefore appears that these mutual dependencies of $\beta$\textit{-catenin} and
members of the \textit{wnt}-signaling family together with its inhibitors
form a \textit{wnt}-related signaling gradient over the full body. 

Our variable $w$ represents this long-range  \emph{wnt}-signaling family. Although it is often argued that a corresponding signal, produced by head cells and 
degraded by tail cells, play a role in regeneration, we do not incorporate an additional variable modeling such a signal since at present understanding the additional information in such a signal would be equivalent to $1-w$ and therefore not contribute in  a mathematically essential way.

\paragraph{Reaction and diffusion.}
We model the space-time evolution of these six concentrations 
$U=(s,h,d,u_h,u_d,w)$ via a reaction-diffusion system
\begin{equation}\label{e:rd}
 \partial_t U = \mathcal{D} U + \mathcal{F}(U),\qquad \mathcal{D}=\mathrm{diags}\,(D_s,D_h,D_d,D_{u_h},D_{u_d},D_w),\quad  \mathcal{F}=(F_s,F_h,F_d,F_{u_h},F_{u_d},F_w). 
\end{equation}
Boundary conditions are explained in Section \ref{s:3.2}. First, we 
discuss the specific forms of the production rates $F_j$, $j\in \{s,h,d,u_h,u_d,w\}$. 

\paragraph{Stem cell production and differentiation rates.}
We assume that stem cells proliferate, undergo apoptosis  \cite{Baguna}, and differentiate irreversibly into other cell types guided by positional control 
genes, that is,  the head and tail related signals,
\begin{equation}
\partial_t s=D_s\partial_{xx}s+p_s(s)-p_hu_hs-p_du_ds, \label{Cell_model_s}
\end{equation}
where $p_s(s)=p_s \frac{s}{1+s} - \eta_s s$ encodes saturated proliferation and apoptosis. 
The localized signals  $u_h, u_d$ trigger differentiation into the 
associated cell lines 
and are subsequently produced by 
these cells. This system acts during both, normal tissue turnover and regeneration
\cite{reddien2004fundamentals}.

\paragraph{Head and tail cells.}
Head and tail cells do not proliferate but result from 
differentiation of stem cells and  undergo apoptosis, leading to 
\begin{align}
\partial_t h&=D_h\partial_{xx}h+p_hu_hs-\eta_hh, \label{Cell_model_h}\\
\partial_t d&=D_d\partial_{xx}d+p_du_ds-\eta_dd \label{Cell_model_d}.
\end{align}

\paragraph{Signal production and depletion rates.}

The differentiated cells produce the signal associated with their cell type 
and  degrade signals associated with other cell types, 
\begin{align}
 \partial_t u_h &= D_{u_h} \partial_{xx}u_h + 
h^2 (r_0  - r_1 u_h) - r_2 u_h d - r_3 u_h,\label{Cell_model_uh}\\
 \partial_t u_d &= D_{u_d} \partial_{xx}u_d + 
d^2 (r_0  - r_1 u_d) - r_2 u_d h - r_3 u_d, \label{Cell_model_ud}\\
 \partial_t w &= D_w \partial_{xx}w - p_w h w
+ p_w d (1-w) \  .   
\end{align}
Specifically, $r_0$ and $r_1$ measure 
saturated  production of cell-type related short-range signals, 
$r_2$  degradation of short-range signals in reactions with
the respective other cell type, and $r_3$ otherwise induced losses of the
signals. 
The quadratic dependence of production of $u_h$ and $u_d$ on $h$ and $d$, 
respectively, is relevant, since replacing this reaction by first-order 
kinetics, that is,  linear in $h,d$, would not result in tristability and spontaneous growth of head and tail regions; see \eqref{e:equdh} in our analysis and model reduction, Section \ref{s:4.1}. 

The kinetics of the \textit{wnt}-signaling complex $w$ are motivated by simple degradation via head cells and saturated production via tail cells. 
We have used the same parameter $p_w$ for degradation and production rates. We tested differing rates with qualitatively similar outcomes. 
Reducing for instance to a smaller saturated production rate, the tail simply
regenerates more slowly, that is, only after the head is formed. Grafting experiments seemed insensitive to changes in rates. Growth experiments fail only when the saturated production rate is too small to compensate for the effect of dilution. 

\subsection{Dynamic boundary conditions and \textit{wnt}-signaling gradients}\label{s:3.2}

Tissue loss, for instance from cutting, leads to a local peak of stem cell differentiation 
\cite{wenemoser2010planarian}. This reaction, which is specific 
to wounds with loss of tissue is not completely understood \cite{REG2:REG256}, but appears to be important for regeneration. 

We choose dynamic (or Wentzel) boundary conditions to model such strong 
effects near cutting sites during wound healing. These somewhat non-standard  boundary conditions supplement the reaction-diffusion system \eqref{e:rd} with inhomegeneous Dirichlet boundary conditions and evolution equations for the time dependent Dirichlet data,
\begin{align}
 \partial_t U &= \mathcal{D} U_{xx}+\mathcal{F}(U),\qquad  U_{x=\pm L}=U_\pm\\
 \frac{\rmd}{\rmd t}U_\pm &=-\frac{1}{\gamma} D \left.\partial_\nu U\right|_{x=\pm L}+\mathcal{F}(U_\pm)+\mathcal{B}.
\end{align}
We think of the concentrations $U_\pm$ as measures for spatially constant concentrations in a \emph{boundary compartment}. Within this boundary compartment, 
close to the respective body edges, the same reactions $\mathcal{F}$ as in the 
central body parts/bulk take place, and diffusive flux terms $ -\frac{1}{\gamma} D \left.\partial_\nu U\right|_{x=\pm L}$ guarantee mass conservation up to production terms,  as can be seen from the short calculation
\[
\frac{d}{dt}\left(\int U +\sum_\pm \gamma U_\pm\right)=\int\mathcal{F}(U) +\sum_\pm \left(\left. \mathcal{D}\partial_\nu U\right|_{\pm L}+\gamma \frac{d}{dt}U_\pm\right) = \int \mathcal{F}(U) + \gamma\sum_\pm \left(\mathcal{F}(U_\pm)+\mathcal{B}\right),
\]
where $\gamma$ is the size of the boundary compartment (or the body edge), measured in 
unit lengths. 
The key difference between these boundary compartments and the bulk 
of the domain is the presence of the terms $\mathcal{B}$ that we shall describe, below. 

The main rationale for dynamic boundary conditions is the need to model dynamics in a thin but finite-size boundary compartment close to the body edges, 
where kinetics are slightly different from the planarian trunk. 
We assume that the boundary compartment is comparatively small such that concentrations of all cells and signals are constant inside, that is, $U_\pm$ does not depend on space. 

Dynamic boundary conditions can be compared to nonlinear fluxes. Formally, 
letting the size of the boundary compartments $\gamma$ tend to 0 and assuming 
at the same time rapid reactions $\mathcal{N}_1$ within these boundary 
compartments 
\[
\frac{\rmd}{\rmd t}U_\pm =-\frac{1}{\gamma} D \left.\partial_\nu U\right|_{x=\pm L}+\mathcal{N},\qquad 
\mathcal{N}=\frac{1}{\gamma} \mathcal{N}_1+\mathcal{N}_2,
\]
one finds in the limit $\gamma=0$ the mixed boundary condition $D\partial_\nu U = \mathcal{N}_1(U)$. We shall see later, in the simplified example of a reduced two-species model in Section \ref{s:redrob}, that such a limit to ``instantaneous'' boundary 
conditions does not appear to be feasible in our modeling context. Beyond this apparent mathematical necessity, there 
appears to be biological data suggesting the presence of such regions close
to the body edges, which separate head- and tail regions from the trunk, and 
the concept of pole regions is thought to play a fundamental role in 
regeneration, see \cite{bookrink}. 

Numerical discretization provides yet a different way to rationalize these dynamic boundary conditions. Representing the main body by grid points $U_1,\ldots,U_N$, each carrying mass $hU_j$ where $h$ is the grid size, we would simply add extra grid points $U_0=U_-$ and $U_{N+1}=U_+$, which carry mass $\gamma U_j$, $j=0,N+1$. This can now be thought of as an inhomogeneous spatial grid, imposing no-flux boundary conditions on the thus extended domain. Clearly, letting the size $\gamma$ of the boundary compartment decrease to the order of the grid scale $h$, we loose the concept of dynamic boundary conditions and should interpret the additional reaction terms $\mathcal{B}$ as nonlinear fluxes. We are not aware of a systematic analysis of such limits, connecting discretization, nonlinear fluxes, and dynamic boundary conditions.

\paragraph{Sensitivity to \textit{wnt}-signaling gradients}

The additional terms $\mathcal{B}$ in the boundary dynamics model the strong local peak of differentiation
of stem cells at wounds inside a formed blastema. In our model, this occurs in a region at the end of the interval considered. The differentiation process appears to be guided by \emph{wnt}-signaling  \cite{adell2010gradients} and preserves polarity. Cutting experiments, where body fragments from quite different regions of a planarian regenerate while preserving polarity suggest that absolute values of \emph{wnt}-signaling do not provide the relevant information. In light of these experiments, and in order to restore polarity from thin fragments 
regardless of specific location of the fragment inside the animal, it therefore seems necessary that stem cells detect gradients of \textit{wnt}-related signaling and differentiate accordingly. Indeed, inspecting the \emph{normal derivative} of $w$ at the boundary after cutting in Figure \ref{f:schematics}, one readily sees that the sign of the normal derivative gives clues as to whether differentiation towards head or tail cells should occur. 

Sensing of the gradient of \textit{wnt}-signaling can be accomplished by 
stem cells in many ways, for instance during their (directed) movement towards 
the wounding site. During their movement they may measure the signal at 
different time steps or 
different locations, see \cite{AltW}, \cite{Devreotes}. We do not attempt to model details of 
this sensing process and simply include a lumped reaction term for the differentiation of stem cells that depends sharply on the sign of the normal derivative of $w$. 

%

\begin{figure}
\centering
\includegraphics[width=0.8\textwidth]{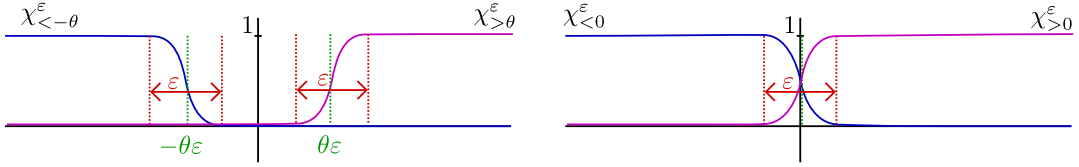}
\caption{
Schematic plot of the indicator functions $\chi$ that detect positive and negative values of the gradient, respectively, with offset $\theta$ and sensing thresholds $\eps$.}\label{f:chi}
\end{figure}

Specifically, we include proliferation and terms for differentiation into the dynamics of $s_\pm,h_\pm,$ and $d_\pm$, of the form
\begin{align}
\frac{\rmd}{\rmd t}s_\pm&=-D_s \left.\partial_\nu s\right|_{x=\pm L}+F_s-\Psi_h^\pm s-\Psi_d^\pm s,\\
\frac{\rmd}{\rmd t}h_\pm&=-D_h \left.\partial_\nu h\right|_{x=\pm L}+F_h+\Psi_h^\pm s,\\
\frac{\rmd}{\rmd t}d_\pm&=-D_d \left.\partial_\nu d\right|_{x=\pm L}+F_d+\Psi_d^\pm s. \ 
\end{align}
The functions $\Psi_{h/d}^\pm$ select differentiation of stem cells 
specifically in the boundary compartments, triggered by the lack of either head or 
tail cells, and directed according to the sign of $\partial_\nu w$,
\begin{align}
 \Psi_{h/d}^\pm&=\Psi_{h/d}(h,d,\partial_\nu w)|_{x=\pm L},\\
 \Psi_h(h,d,\partial_\nu w)&=\tau(1-h)(1-d) \chi_{>\theta}^\eps(\partial_\nu w),\\
 \Psi_w(h,d,\partial_\nu w)&=\tau(1-h)(1-d) \chi_{<-\theta}^\eps(\partial_\nu w).
\end{align}
Here, $\tau$ is the rate of this mechanism and $\chi_j^\eps$ are smooth versions of the characteristic function given in \eqref{e:char}; see Figure \ref{f:chi} for a schematic representation of the indicator functions $\chi$. 

%
%
%
%
%

\subsection{Cutting, grafting, and growth}\label{s:3.3}
We present numerical simulations of our mathematical model illustrating 
homeostasis, cutting, grafting, and growth, confirming and expanding on the schematic representation in Figure \ref{f:schematics}.


\paragraph{Parameter values: diffusivities and rate constants.}
We chose parameter values representing roughly expected orders of magnitude 
within the system. Moderate changes of parameters do not have a significant effect on 
the outcome of simulations but we discuss some notable exceptions, below. The following table shows
 default parameter values, used unless noted otherwise. 
\begin{table}
\centering
\begin{tabular}{p{.8cm} p{3.3cm}  p{.8cm} p{3.3cm}   p{.8cm} p{1.2cm} }
\toprule
$D_s$      &  $1$        & $p_s$ & $100$      & $\eta_s$  & 100   \\
 $D_h$     &  $10^{-3}$  & $p_h$ & $1$       & $\eta_h$   & $1$      \\ 
 $D_d$     &  $10^{-3}$  & $p_d$ & $1$       & $\eta_d$   & $1$      \\ 
 $D_{u_h}$ &  $10^{-2}$  & $p_w$ & $10$      & $\tau$     & $0.5$      \\ 
$D_{u_d}$  &  $10^{-2}$  & $r_0$ & $18$     &  $\theta$   & $3$   \\ 
$D_w$      &  $1$        & $r_1$ & $12$     &  $\eps$    &  $2\cdot10^{-3} $    \\      
$\gamma$   & $0.3$       & $r_2$ & $6$     &          &   \\ 
$L$        &  $10$       & $r_3$ & $6$     &            &     \\ 
\bottomrule                                 
\end{tabular}
\caption{Parameter values used in the numerical simulations throughout, unless noted otherwise.}
\label{table:pars}
\end{table}
Random motion of stem cells is of order one as they move fairly freely 
through the body. Recall that we do not model directed motion of stem cells but recognize that such mechanisms may be necessary to accomplish gradient sensing in a robust fashion. Random motion of head 
and tail cells is very slow. Similarly, local signals $u_{h/d}$ diffuse slowly, while the long-range \textit{wnt}-related signal has a diffusion constant of order 1. We work on domains of length $10$  and attribute a mass fraction $\gamma=0.3$ to the boundary, unless otherwise indicated. We assume fast proliferation of stem cells $p_s\gg 1$ and fast signaling dynamics relative to cell differentiation, $r_j\gg 1$. Cell differentiation $p_{h/d}$ and apoptosis $\eta_{h/d}$ as well as  differentiation at the tissue edges during wound healing $\tau$ occur on a time scale of order 1. The production rate of \textit{wnt}-related signals
$p_w\gg 1$ is comparatively fast. The constant $\eps$  is small, approximating discontinuous switching in the kinetics, the constant $\theta$  measuring minimal detectable strength of the gradient at the body/wound edge in $\eps$-units is set to $3$. 
We comment below on some of the effects of changing parameter values but did not attempt a systematic study of parameter space.

\paragraph{Numerical implementation.}

We implemented the dynamic boundary conditions as time-dependent Dirichlet conditions. We then solved the system with grid spacing $dx=0.01$ and time stepping $5\cdot 10^{-4}$ using a semi-implicit Euler method. We found little changes from refining spatio-temporal grids and also used  \textsc{Matlab's} stiff solver \textsc{ode15s} for comparison with good agreement.

\paragraph{Homeostasis.}

We obtained equilibrium profiles starting with initial conditions that 
represent head and tail cells in the boundary compartments at the body edges, 
a uniform distribution of stem cells throughout the trunk, and a 
uniform \emph{wnt}-signaling gradient. Solving the initial-value problem for a short time until, we found that concentrations approached constants in time.

\begin{figure}[h!]
\includegraphics[width=0.27\textwidth]{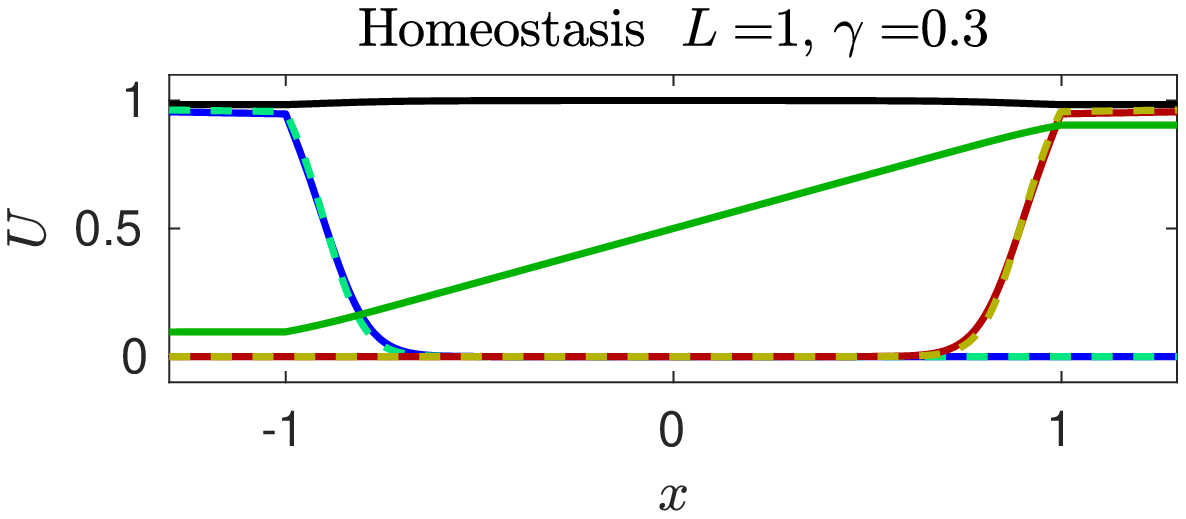}\hfill
\includegraphics[width=0.27\textwidth]{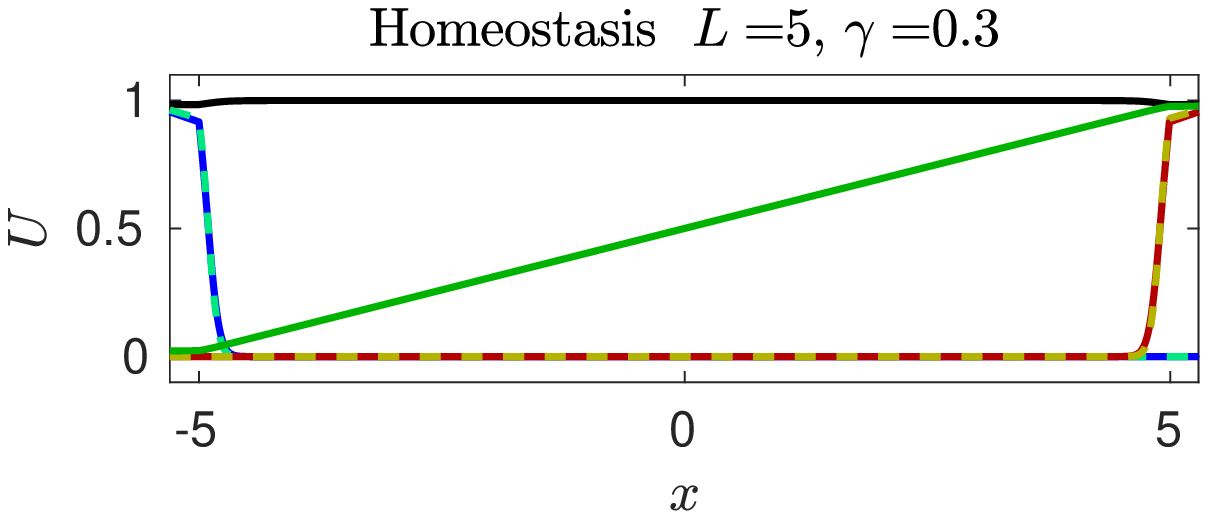}\hfill 
\includegraphics[width=0.27\textwidth]{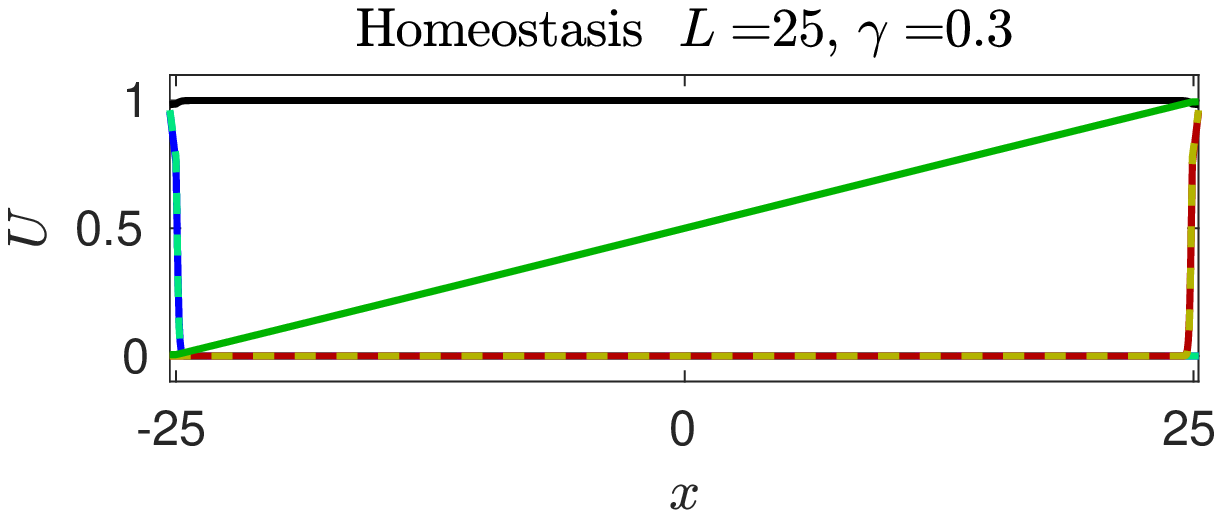}\hfill 
\includegraphics[width=0.129\textwidth]{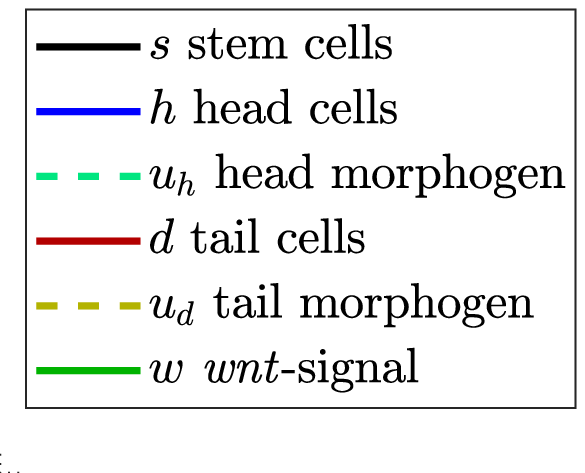}
 \caption{Equilibrium profiles showing a linear \textit{wnt}-signaling profile, head and tail cells concentrated near boundaries, stem cell concentrations with small deviations from constant, and chemical signals closely following head and tail-cell concentrations, respectively. Note the different scales for $x$,
which represent different body sizes of planarians. Homeostasis and subsequent cutting experiments are illustrated in the supplementary materials \textsc{cutting\_sequel.mp4}.}\label{f:homeo}
\end{figure}

Specifically, we used initial conditions
\begin{eqnarray*}
h_0(x)&\equiv 0, &h_{-,0}=1,\ h_{+,0}=0\\
d_0(x)&\equiv 0, &d_{-,0}=0,\ d_{+,0}=1\\
s_0(x)&\equiv 1, &s_{-,0}=1,\ s_{+,0}=1\\
w_0(x)&= \frac{x+L}{2L},& w_{-,0}=0,\ w_{+,0}=1,
\end{eqnarray*}
and let $u_h$ and $u_d$ equal $h$ and $d$, respectively, for the initial data. The results, schematically shown in Figure \ref{f:schematics}, are displayed in Figure \ref{f:homeo} and illustrate the stationary profiles of healthy organisms
of different body size. 
We found that the linear concentration profile in the \textit{wnt}-signal is quite robust under dramatic changes in the domain size. Fixing the width of the boundary compartment $\gamma$ and varying $L$ we found robust homeostasis between $L=0.005$ (!) and $L=40$. For very small $L$, the total variation of  the \textit{wnt}-signal  $w$ decreases: $w$ stays bounded away from $0$ and $1$, and the gradient of $w$ remains bounded  as $L$ tends to zero. In this regime, the signals $u_h$ and $u_d$ follow the concentrations $h$ and $d$, respectively, less closely, being much smaller in amplitude. On the other hand, for very large $L$, the equilibrium state is sensitive to small fluctuations since the \textit{wnt}-signaling gradient is 
small, within the  interval $[-\theta \cdot \eps,\theta \cdot \eps]$ where 
detection of the gradient is no longer decisive.
Increasing the size of the boundary compartment $\gamma$ helped stabilize dynamics generally and we found robust homeostasis for domain sizes 
$1, \ldots , 50$ and  boundary compartments of size $\gamma=0.1, \ldots  , 15$. 
Very small sizes of the boundary compartment, such as $\gamma=0.015$, $L=5$, were not able to sustain a head-trunk-tail profile, consistent with our discussion in Section \ref{s:4}, below. Whenever we saw homeostasis, we also tested robustness against small amplitude perturbations and found recovery within expected limits. A typical dangerous perturbation would for instance attempt to alter the sign of the \textit{wnt}-signaling gradient near the boundary. We discuss perturbations more in line with some experiments, below. 

\begin{figure}[h!]
\includegraphics[width=0.295\textwidth]{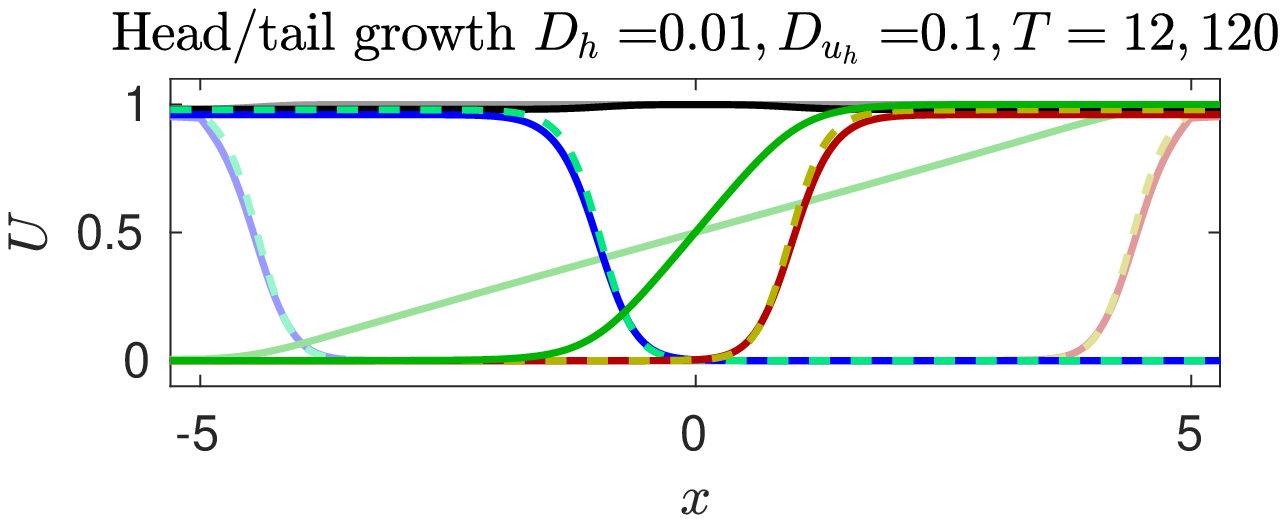}\hfill
\includegraphics[width=0.275\textwidth]{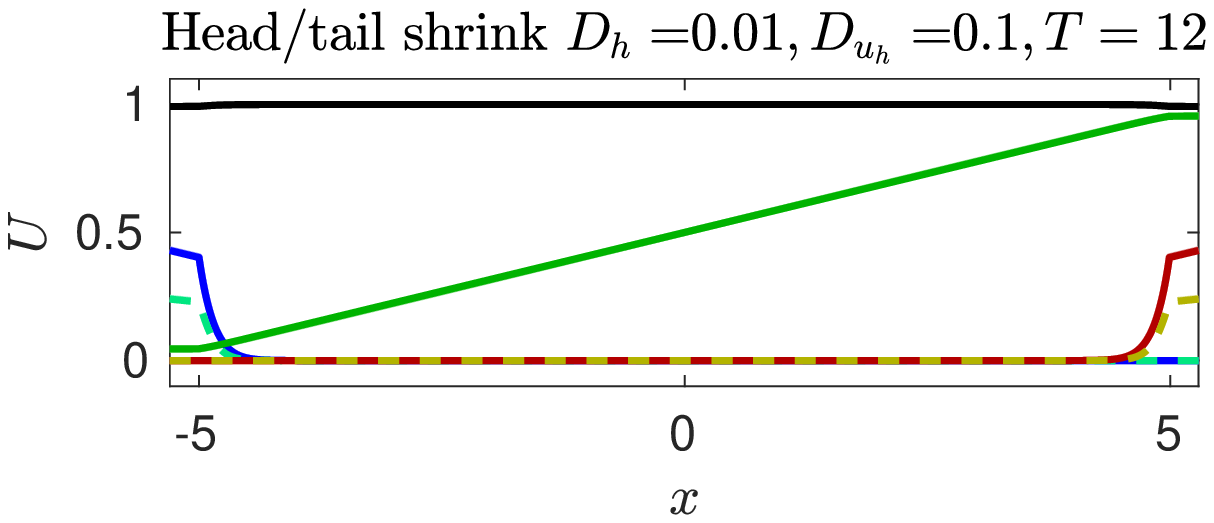}\hfill 
\includegraphics[width=0.275\textwidth]{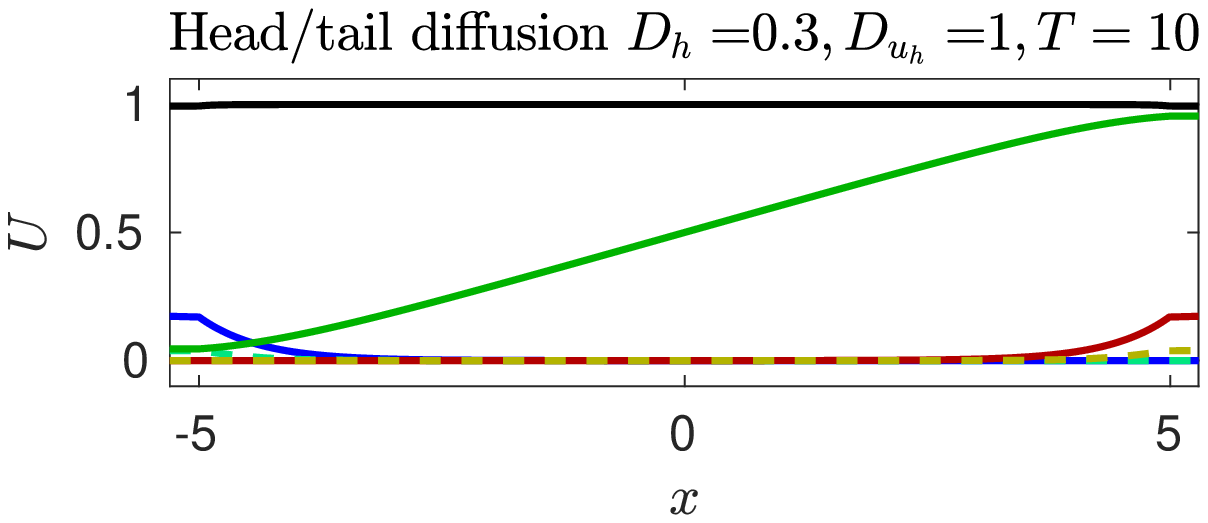}\hfill 
\includegraphics[width=0.129\textwidth]{homeolegend.eps}
 \caption{Growing and shrinking of head  and tail cell regions (left/center) 
with $r_0=16$, $r_3=4$ (left) and  $r_0=18$, $r_3=10$ (center); 
initial conditions: $s\equiv 1$, linear $w$-gradients, no head or tail cells. 
Expansion of head/tail regions shown on the left with snapshots at $T=12$ 
(opaque) and $T=120$ (solid);  shrinking terminates and homeostasis is reached at $T=12$.  
 Right, influence of strong random motion and diffusion on homeostasis. 
Throughout, parameters are as in Table \ref{table:pars} unless noted otherwise. Compare also the visualization in the supplementary materials \textsc{cutting\_sequel.mp4}.}\label{f:homeo2}
\end{figure}

The localization of the regions occupied by head and tail cells depends first on the strength of random motion $D_d,D_h$ of $d$ and $h$ and diffusion rates of 
their associated morphogens; 
see Figure  \ref{f:homeo2}. On the other hand, changing production and degradation rates $r_0,r_1,r_2,r_3$ for $u_{h/d}$, 
one can trigger a spontaneous expansion of the region of head and tail cells, where now the rate of expansion depends on these rates 
and the diffusivities.

\begin{figure}[h!]
\includegraphics[width=0.27\textwidth]{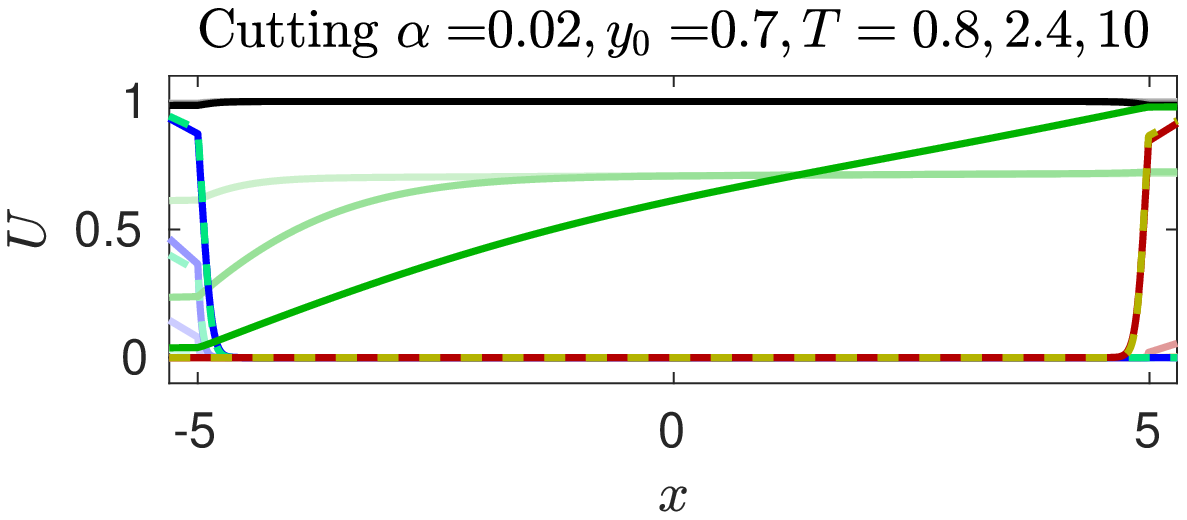}\hfill
\includegraphics[width=0.27\textwidth]{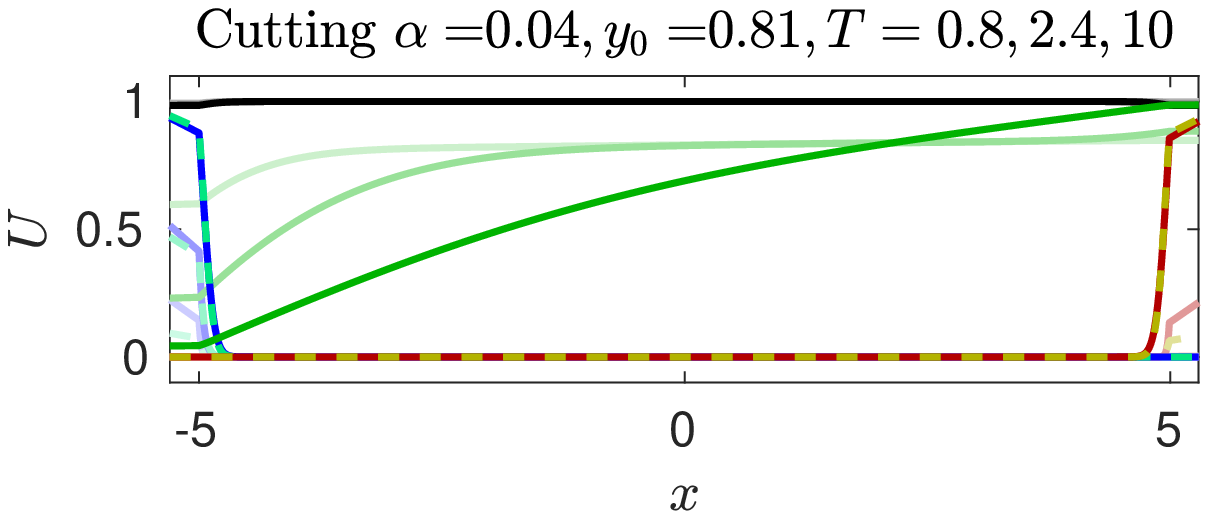}\hfill 
\includegraphics[width=0.27\textwidth]{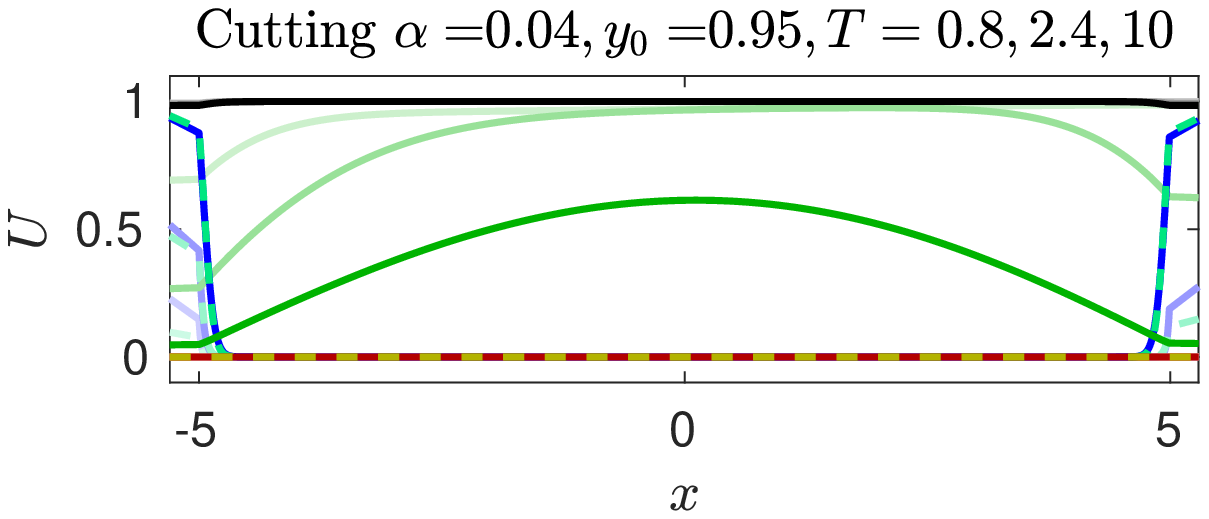}\hfill 
\includegraphics[width=0.129\textwidth]{homeolegend.eps}
 \caption{Regeneration after cutting experiments: cutting 1/50th from trunk 
region $y_0=0.7$ (left), 1/25th close to tail $y_0=0.815$ (center) and from 
tail $y=0.95$ (right); see \eqref{e:alpha} for initial conditions. Note the 
failure to regenerate through the emergence of heads on both sides and the 
loss of the \textit{wnt}-signaling gradient on the right, for a cut very 
close to the tail region. Throughout, parameters are as in Table 
\ref{table:pars}; opaque curves show concentrations at earlier snapshots as 
listed in the title.}\label{f:cutting}
\end{figure}

\paragraph{Cutting.}
Relating to experiments where a strip of the trunk of a planarian is cut out, 
we now study initial conditions that arise when using part of the homeostatic 
state in the central region of the domain, that is, 
\begin{equation}\label{e:alpha}
 w=\frac{\alpha (x+L)}{2L}+y_0, \quad s\equiv 1, \quad h\equiv d\equiv u_h\equiv u_d\equiv 0,
\end{equation}
where $\alpha$ represents the fraction of the fragment cut out, and 
$y_0\in [0,1]$ the $w$-concentration at the left edge of the cut fragment, corresponding to a cutting location $x_0=L(2y_0-1)\in[-L,L]$. In other words, 
$\alpha=0.02$ corresponds to a cut of $2\%$ of body length and $y_0=0.75$ corresponds to a cut at three quarters of the body lengths distance from the head. Boundary data is chosen compatible $U_\pm=U(\pm L)$ at time 0. Figure \ref{f:cutting} demonstrates recovery from cuts of 2\% and 4\% of body length. Recovery is more sensitive if segments are cut out from tail or head regions. As expected, recovery depends on the sensitivity $\eps^{-1}$ 
and the threshold $\theta$. Smaller values of $\eps$ and $\theta$ increase sensitivity and enhance recovery. On the other hand, small values of $\theta$ and $\eps$ also enhance vulnerability to noise and its effect on \textit{wnt}-signaling gradients near the boundary. Smaller mass fractions $\gamma\lesssim 0.03$ 
in the boundary compartments for a domain with $L=5$ prevent recovery even for medium-size cuts, to the same extent as homeostasis breaks down. Generally, for cuts near the tail, the head regenerates first, and vice-versa. Throughout the simulations, polarity is consistently preserved. 

In our numerical simulations, one observes an initial strong burst in 
differentiation triggered by the boundary sensing $\Psi_{h/d}^\pm$. This is 
compensated for by the strong proliferation of stem cells $p_s$. Simulations 
still show a resulting decrease in the stem cell population at the boundary, $s_\pm < \frac{1}{2L}\int s$, which can be much more pronounced for different parameter values such as smaller $p_s$ or larger $\tau$. This reflects and quantifies in our model the experimentally observed strong proliferation of stem cells during wound healing.

\begin{figure}[h!]
\includegraphics[width=0.27\textwidth]{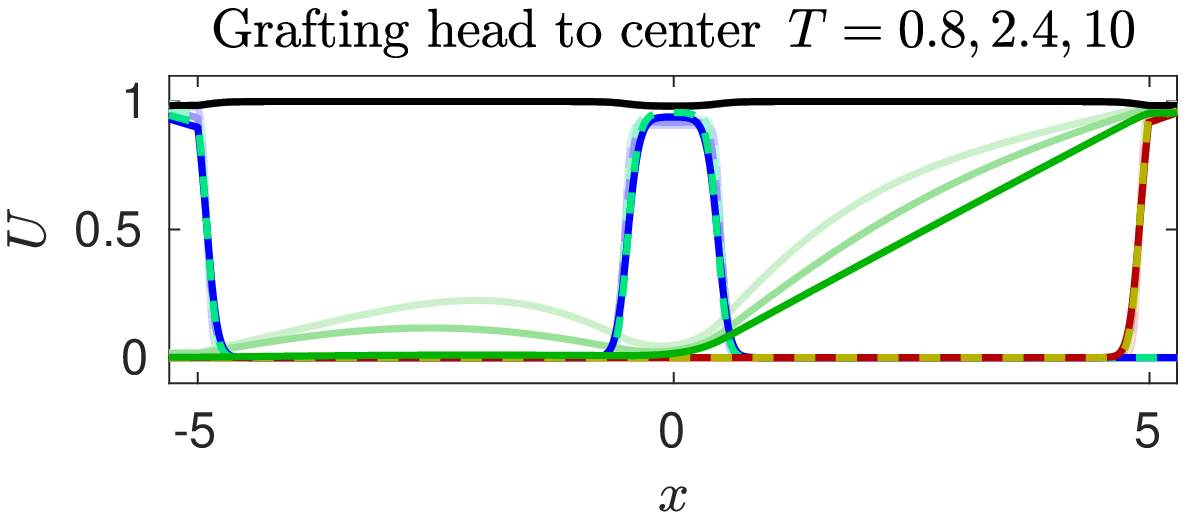}\hfill
\includegraphics[width=0.27\textwidth]{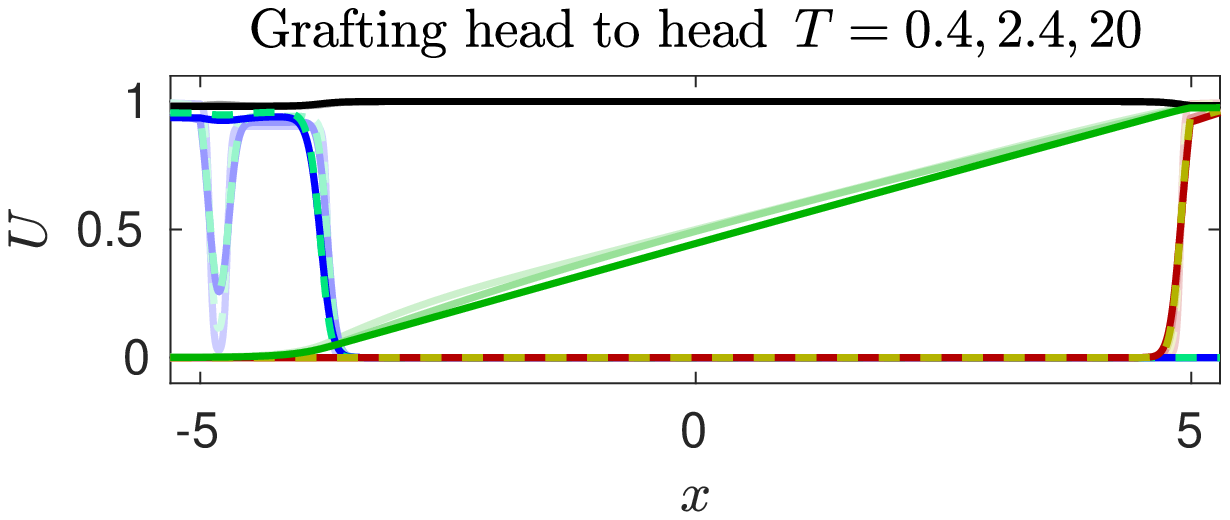}\hfill 
\includegraphics[width=0.27\textwidth]{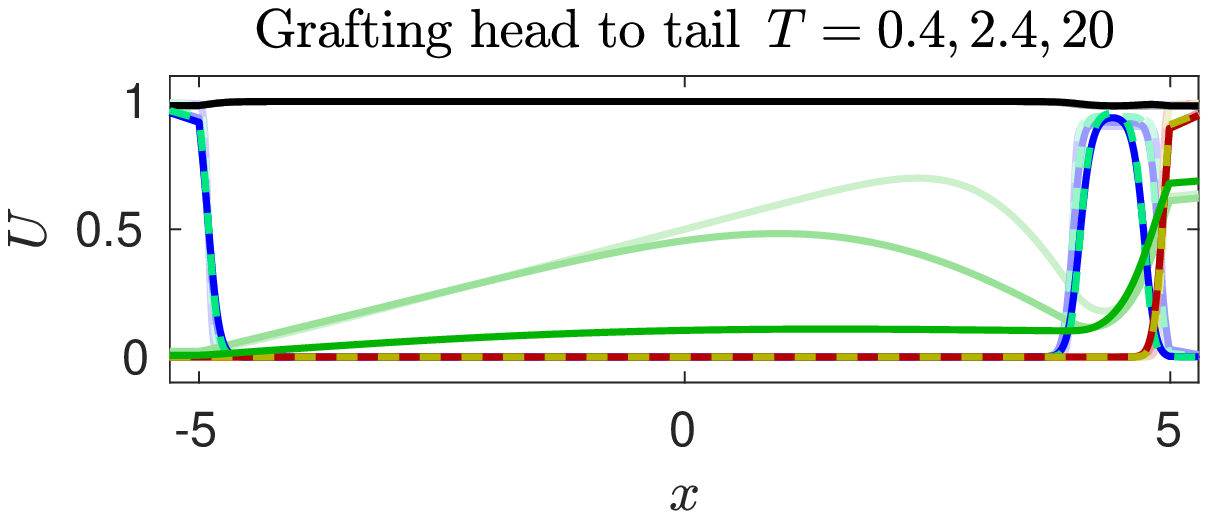}\hfill 
\includegraphics[width=0.129\textwidth]{homeolegend.eps}
 \caption{Regeneration after grafting head tissue into center of trunk (left), head region (center), and tail region (right). Throughout, parameters are as in Table \ref{table:pars}; opaque curves show concentrations at earlier snapshots as listed in the title. Note in particular the persistence of the grafted head in the left and right panels versus the merging (and eventual vanishing) of the graft in the center panel.  Compare also the visualization in the supplementary materials \textsc{grafting.mp4}.}\label{f:grafting}
\end{figure}

\paragraph{Grafting.}
In this third set of numerical experiments, we study initial conditions where head cells are inserted into a healthy planarian. We therefore use the results 
of the homeostasis simulation as initial condition and then change values in a 
region of size 10\% of the length of the organism. In this region, we increase  the concentration of head cells $h$ and the associated morphogen $u_h$ to $1$ and eliminate \emph{wnt}-signaling, that is, set $w\equiv 0$. We observe that the 
head region survives and changes the profile of \textit{wnt}-signal concentrations. Grafting head cells of a donor close to the head region of 
the host leads to merging 
of the two head regions. Grafting donor head cells near the tail of the host can preserve the tail, unless tail cells are significantly destroyed during grafting. Figure \ref{f:grafting} shows numerical experiments corresponding to these three scenarios. We think of these different outcomes, for instance the possibility of merging of heads vs persistence of the new head as reflecting the dichotomy 
seen in experiments where an additional head may grow out of a graft or the graft may just disappear; see the discussion in Section \ref{s:2} and Figure \ref{f:schematics}.

We caution here that these numerical experiments rely to some extent on astute choices of production rates for head and tail cells and the associated morphogen, as well as the choice of diffusivity. In our experiments, the boundary between head and trunk cells is nearly stationary and sharply localized. Roughly speaking, an open set of parameter values will lead to expanding head (or tail) regions, a complementary open set will lead to shrinking head (or tail) regions. At the boundary of these parameter regions, head and tail regions will remain nearly stationary, subject only to a slow coarsening interaction  as seen in the head-to-head grafting. Stronger diffusivities will enhance both speed of growth and shrinking as well as the slow coarsening. Our choices of parameters are close to 
the critical values, where head and tail regions neither shrink nor expand. We emphasize that for parameters where head- or tail regions shrink, grafted regions will slowly disappear, but the head- and tail regions near the boundary will persist due to the \textit{wnt}-related gradient triggered differentiation of stem cells. 

We did not explore the phenomena of growth and shrinking of planarians in 
much more detail than in the next paragraph,  
since our model so far does not include any well-motivated mechanisms that would regulate the size of organs.

We note that regions between two heads (or between two tails) do not generate 
a \textit{wnt}-signaling gradient. One would therefore predict that 
secondary cutting 
experiments with tissue from such regions after grafting would not be able to 
regenerate consistently and not preserve polarity. 

\begin{figure}[h!]
\includegraphics[width=0.27\textwidth]{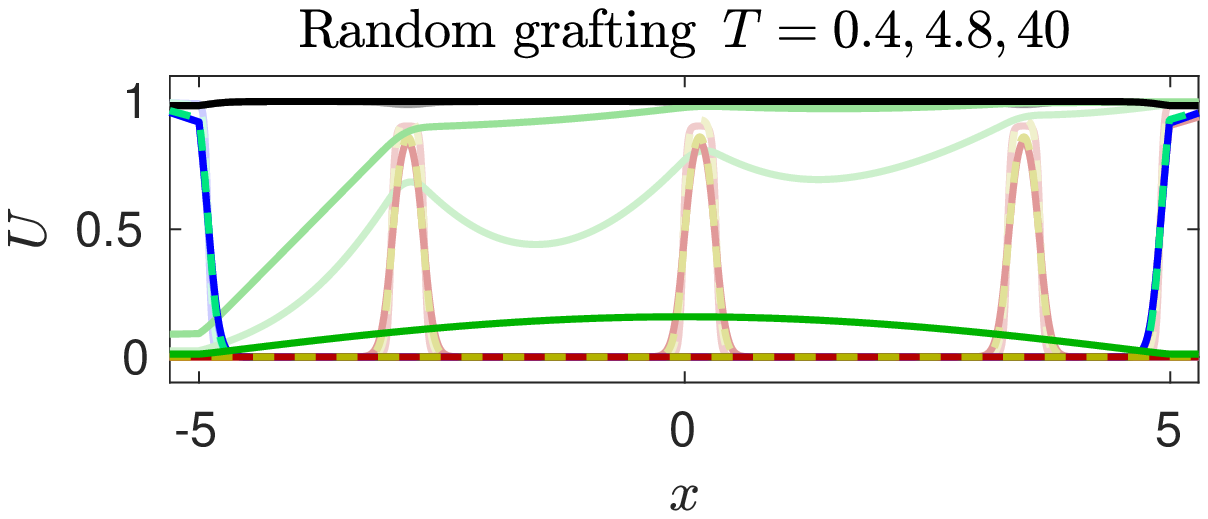}\hfill
\includegraphics[width=0.27\textwidth]{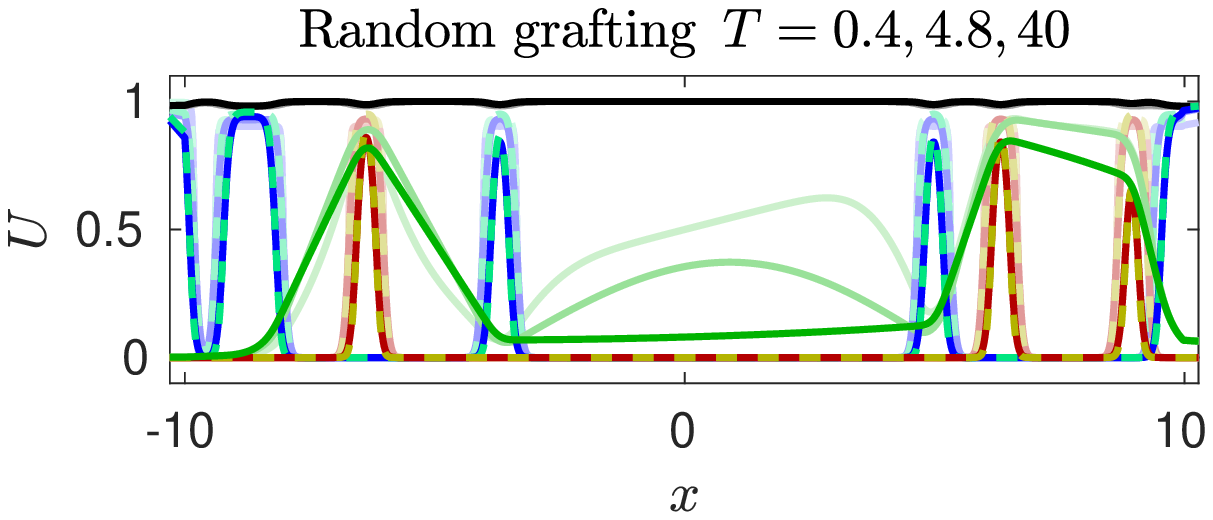}\hfill 
\includegraphics[width=0.27\textwidth]{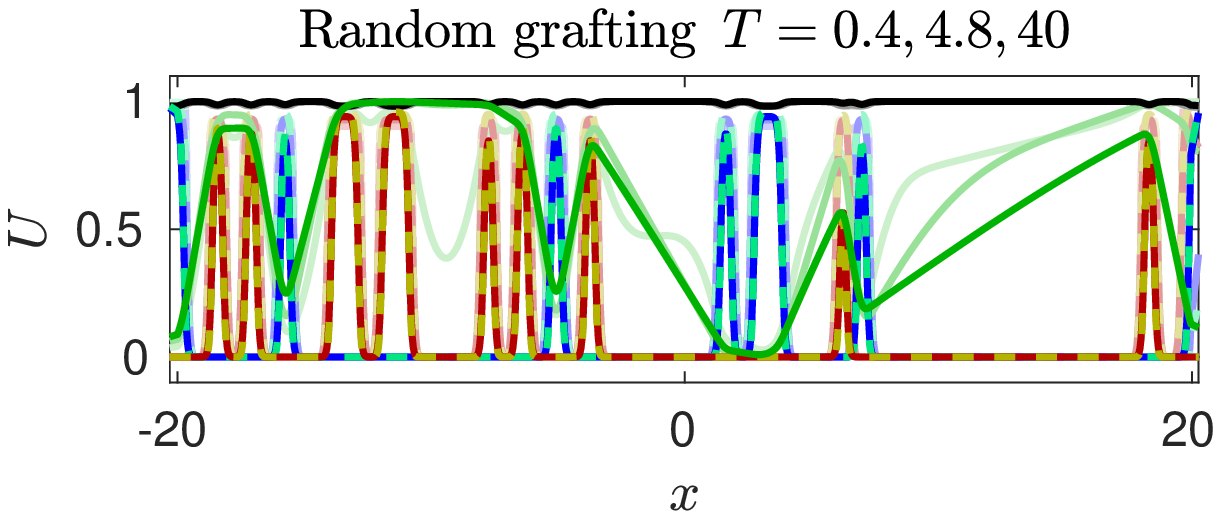}\hfill 
\includegraphics[width=0.129\textwidth]{homeolegend.eps}
 \caption{Regeneration after grafting multiple head and tail regions, for different body lengths $L=10,20,40$.  
Throughout,  parameters are as in Table \ref{table:pars}; opaque curves show 
concentrations at earlier snapshots as listed in the title.
Annihilation of all grafts on the left, and multiple persistent regions for larger domains (center and right).}\label{f:graftingrand}
\end{figure}

Mimicking dissociation experiments in hydra, we also explored outcomes of 
random grafting in our mathematical model, see 
Figure \ref{f:graftingrand}. We  inserted head and tail fragments of roughly 3\% of body length randomly at various locations in the simulated
planarian. Whenever these pieces are large enough, they persist as head and tail regions; smaller regions eventually disappear. 

\begin{figure}[h!]
\includegraphics[width=0.27\textwidth]{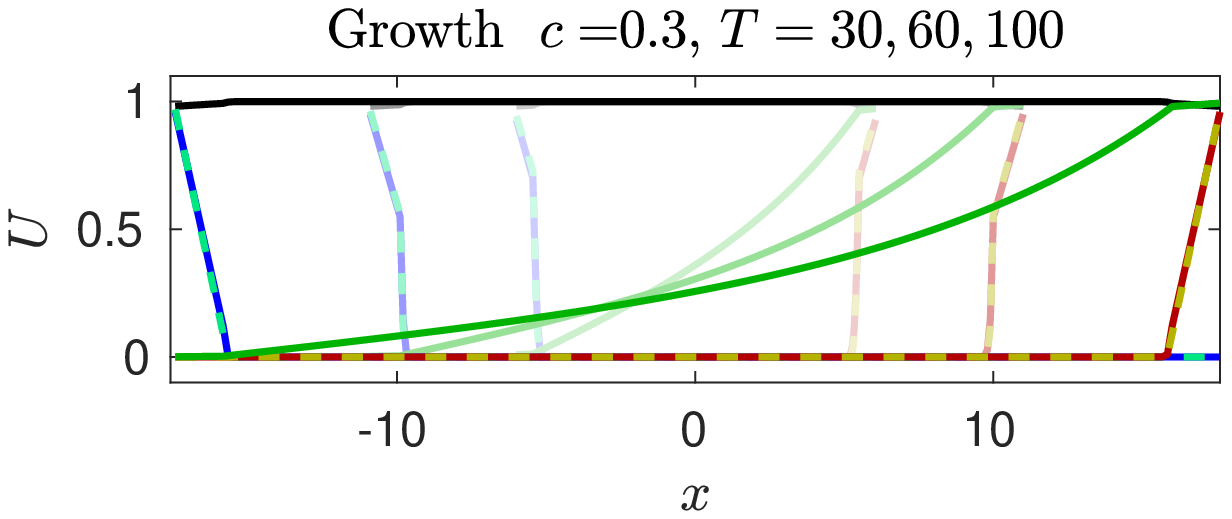}\hfill
\includegraphics[width=0.27\textwidth]{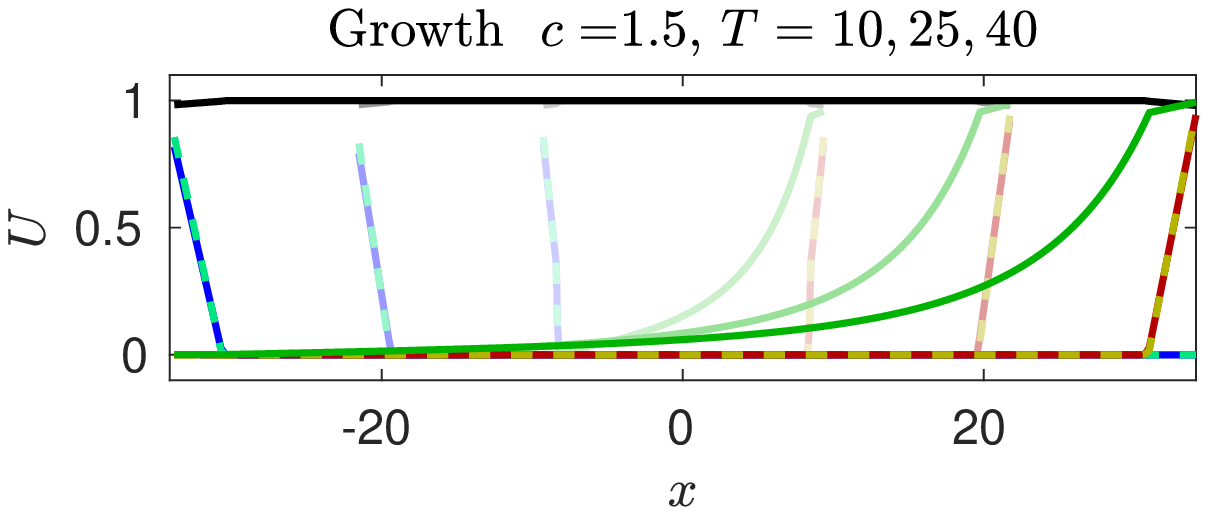}\hfill 
\includegraphics[width=0.27\textwidth]{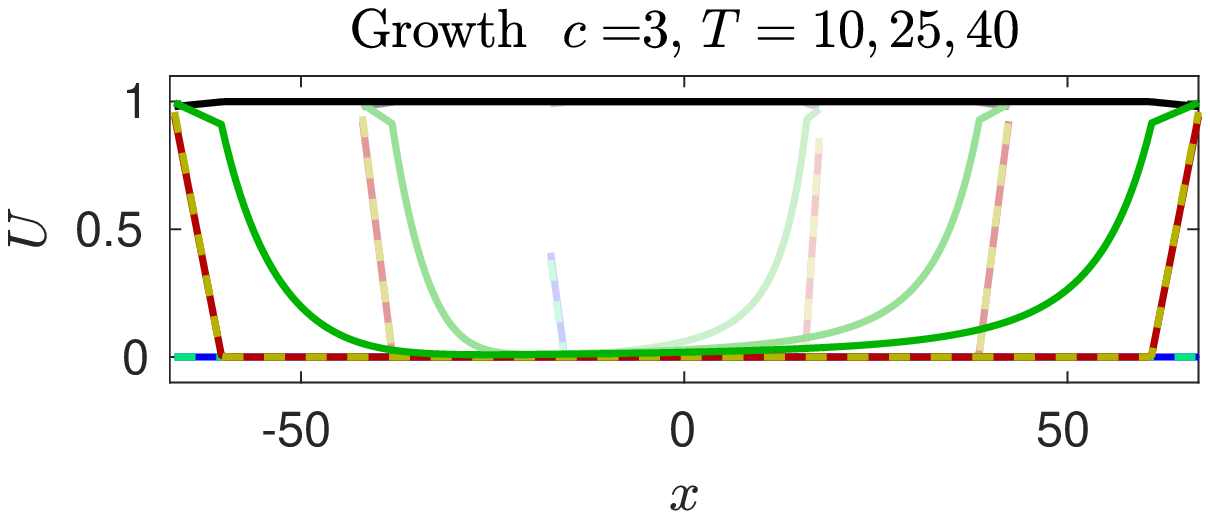}\hfill 
\includegraphics[width=0.129\textwidth]{homeolegend.eps}
 \caption{A healthy planarian under uniform linear growth with speeds $c=0.3,\,1.5,\,3$; see text for more details.  Concentration profiles are plotted in actual coordinates, such that they occupy only part of the final domain at earlier times (faded curves terminate at $\pm (L_0+cT)$). Profiles are well maintained close to homeostasis for slow (left) and moderate (center) growth speeds. For larger speeds (right) dilution of $w$ reduces the overall concentration such that the gradient at the left boundary where $w\sim 0$ falls below the sensing threshold. As a result concentration of head cells decreases on the left boundary and a second tail replaces the head.  }\label{f:growth}
\end{figure}

\paragraph{Growth.}
As a last experiment, we demonstrate that the mechanisms of our model are capable of sustaining robust patterning when the body size  of the
planarian expands. In our simulations, the body size expands uniformly at a constant speed $c=0.3,1.5,3$ and we monitor the concentration profiles; see Figure \ref{f:growth}. More precisely, we assume that tissue locations evolve according $x(t)= \frac{x(0)}{L_0} (L_0+ct)$, where $(-L_0,L_0)$ is the body size at time $t=0$ and $(-L(t),L(t))$, $L(t)=L_0+ct$, is the body size at time $t$.  Mass conservation in the extended domain, in the absence of reaction terms, then forces the dilution  $\partial_tU(t,x)=\rho(t) U(t,x)$, with $\rho(t)=-\frac{c}{L(t)}$. Transforming back to a fixed domain with a coordinate change $x\mapsto x L_0/L(t)$ gives a diffusion equation on $(-L,L)$ with diffusion matrix $\frac{L_0^2}{L(t)^2} D$. In the abstract formulation \eqref{e:rd}, we therefore simply amend the diffusion constant in our model and add a dilution term to model the system on the growing domain,
\begin{equation}\label{e:rdgr}
 \partial_t U = \mathcal{D}(t) U_{xx} + \mathcal{F}(U)+\rho(t) U, \qquad \mathcal{D}(t)=\frac{L_0^2}{L(t)^2}\mathcal{D},\quad \rho(t)=-\frac{c}{L(t)}.
\end{equation}
We chose to preserve the relative size of the boundary compartments such that $\gamma=\gamma(t)=\frac{\gamma(0)}{L_0}L(t)$, which introduces the same dilution term in the equation for the dynamic boundary values $U_\pm$. Fluxes need to be adjusted to the changing diffusion constant $-\frac{1}{\gamma(t)}\mathcal{D}(t)\partial_\nu U$. We refer to \cite{tenbrock} for a more detailed discussion of different growth laws and references to the literature. 

We simulated the resulting system on a fixed grid corresponding to scaled $x$-coordinates but plotted the outcome in the actual unscaled $x$-variables. Parameters were chosen according to Table \ref{table:pars}.
The results are displayed in Figure \ref{f:growth}.
For  slow and moderate growth speeds, the concentration profiles resemble the homeostatic profiles at a given length. We observe failure to maintain key features only for quite rapid expansion (the time scales of $c=3$ corresponds to a doubling of a planarian of size 9 in 3 time units, while cell differentiation happens with rate 1, generating at most 1 unit volume of head or tail cells from stem cells in one time unit). This failure at rapid growth can be attributed to the dilution of the \emph{wnt}-related signal that is not adequately compensated for by production through tail cells and diffusion. As a result, the \textit{wnt}-signaling gradient becomes very small in the head region. As already demonstrated in the case of fixed domains  during homeostasis and cutting, solutions become sensitive to perturbations when \textit{wnt}-signaling gradients approach the sensing limits determined by $\eps$ and $\theta$.

Of course, the externally imposed growth used here is not a completely appropriate representation of the actual growth of planarians while feeding. Conversely, the model can not yet clearly characterize the dynamics under starvation conditions when sizes of planarians shrink by an order of magnitude: the negative dilution term would give an overcrowding of cells with sometimes unstable signaling gradients near the boundary. A more accurate model would relate growth laws and proliferation, which in turn depends on food supply. Nevertheless, the scenario studied here does show the inherent independence of the patterning from body size and its robustness for growing planarians.

\section{Analysis via model reduction}\label{s:4}
We first argue that stem cells and short-range signals can be eliminated in a quasi-steady state approximation in Section \ref{s:4.1}. We reduce the resulting system for $(h,d,w)$ further by considering one order parameter $c$ for the cell type instead of the two concentrations $(h,d)$ in Section \ref{s:4.2} and illustrate how this reduced two-species system captures regeneration after cutting and grafting in Section \ref{s:4.3}. Finally, we focus on the key element of the process, the restoration of the \textit{wnt}-related signal in a further reduced scalar equation.

\subsection{Eliminating stem cells and short-range signals}\label{s:4.1}

A simple analysis of kinetics, supported by observations in numerical simulations, suggests two quick simplifications by reduction. 

\paragraph{Stem cell populations are constant in time and space.} Inspecting the proliferation and differentiation rates of stem cells, one notices that proliferation is much faster than differentiation and equilibrium concentrations depend only marginally on differentiation, that is, on the concentrations of $u_{h/d}$. Setting $p_s(s)=0$, we then obtain equilibrium concentrations $s=(p_s-\eta_s)/\eta_s$, namely, in our choice of parameters,  $s\equiv 1$. 

\paragraph{Short-range signals are tied to cell populations.} 
Assuming in the following that $s\equiv 1$, we notice that the signal production rates $r_j$ are much larger than the cell differentiation and death rates $p_{h/d}$ and $\eta_{h/d}$. Using an adiabatic reduction for the kinetics, one then equilibrates the reaction rates for $u_{h/d}$ and finds $u_{h/d}$ as functions of $h$ and $d$, 
\begin{equation}\label{e:equdh}
u_h=\frac{r_0 h^2}{r_1 h^2+r_2 d + r_3},\qquad u_d= \frac{r_0 d^2}{r_1 d^2+r_2 h + r_3}.
\end{equation}
Similar to the case of stem cell dynamics, the reduction for the kinetics is valid under suitable bounds on gradients, and we shall demonstrate below that such reduced systems capture the key structure of the dynamics quite well. 

In summary, the reduction thus far, substituting the expressions for $u_{h/d}$ into the equations for $h,d,w$, after setting $s\equiv 1$, yields 
the system
\begin{eqnarray}
\partial_t h &=& D_h \partial_{xx} h + p_h \frac{r_0 h^2}{ r_1 h^2 + r_2 d+r_3} -\eta_h h \    , \nonumber \\
\partial_t d &=& D_d \partial_{xx} d + p_d \frac{r_0 d^2}{ r_1 d^2 + r_2 h+r_3} -\eta_d d \   , \label{hdw} \\
\partial_t w &=& D_w \partial_{xx} w  - p_w hw + p_w d (1-w) \   .  \nonumber
\end{eqnarray}
For the boundary data, we find
\begin{eqnarray*}
\frac{d}{dt}h_\pm       &=& -\frac{1}{\gamma}D_h    \partial_\nu h_\pm  
+ p_h \,\frac{r_0 h_\pm^2}{ r_1 h_\pm^2 + r_2 d_\pm+r_3} -\eta_h h_\pm +\Psi_h^\pm,\\
\frac{d}{dt}d_\pm       &=&-\frac{1}{\gamma}D_d    \partial_\nu d_\pm  
+  p_d \, \frac{r_0 d_\pm^2}{ r_1 d_\pm^2 + r_2 h_\pm+r_3} -\eta_d d_\pm+\Psi_d^\pm, \\
\frac{d}{dt}w_\pm       &=& -\frac{1}{\gamma}D_w    \partial_\nu w_\pm    
- p_w h_\pm w_\pm  + p_w d_\pm(1-w_\pm)) \  .
\end{eqnarray*}
where again 
\[
\Psi_h(h,d,\partial_\nu w)=\tau(1-h)(1-d) \chi_{>0}^\varepsilon(\partial_\nu w), \qquad 
 \Psi_w(h,d,\partial_\nu w)=\tau(1-h)(1-d) \chi_{<0}^\varepsilon(\partial_\nu w)\ . 
\]
Simulations of this reduced model are almost indistinguishable from the full model and therefore we do not display the somewhat redundant graphics here. We proceed by a further, more phenomenological simplification. 

\subsection{From cell type to order parameter}\label{s:4.2}
In numerical simulations of our mathematical model, which are related to the 
main biological experiments described above, concentration profiles of $h$ and $d$ are mostly constant, taking on values $(h_*,0)$, $(0,d_*)$, or $(0,0)$, where in fact $h_*=d_*= h_+$, 
\[
h_\pm=\frac{p_h r_0\pm \sqrt{p_h^2 r_0^2 - 4 \eta_h^2 r_1 r_3}}{2 \eta_h r_1}, 
\]
provided that  $p_h^2r_0^2>4\eta_h^2 r_1 r_3$, which we shall assume in the sequel. 
These three states are in fact stable equilibria for the ODE kinetics and for the full PDE \eqref{hdw}
for $h,d$, (due to the equal diffusivites)  and  correspond to head-only, tail-only, and trunk-only states, respectively.  
The equilibria $(h_-,0)$ and $(0,h_-)$ satisfy $0<h_-<h_+$ and are unstable threshold states, separating initial conditions that evolve toward head cells from initial conditions evolving toward trunk-only cells.  Due to the symmetric choice of parameter values, pure head and tail states have equal concentration values here, although this is not necessary in order to capture the 
considered phenomena. 

Regions of constant values of $h$ and $d$ are separated by interfaces (or fronts) that propagate with a speed  determined by the parameter values $r_j$, $\eta_j$, and $p_j$. There are three possible fronts, head-trunk, tail-trunk, and head-tail. 

\begin{figure}
\centering\includegraphics[width=0.6\textwidth]{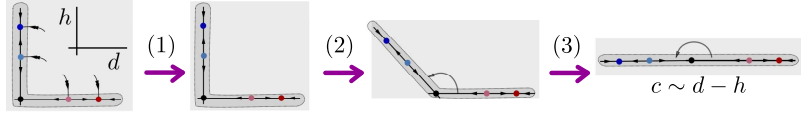}
\caption{Schematic of the reduction from cell type to order parameter. In the kinetics of the head-cell dynamics, the $\omega$-limit set of a large ball consists of the line segments of the coordinate axis between the origin and pure-head and pure-tail state, respectively. Restricting to this $\omega$-limit set (1) gives the joined line segments, which one then (2) bends open to (3) arrive at a straight one-dimensional line segment with dynamics equivalent to the scalar order parameter $c\sim d-h$; see text for a rationale for this reduction including diffusive effects, particularly fronts.}\label{f:threetotwo}
\end{figure}

Fronts between pure head and trunk state can be understood in the $h$-equation
alone, setting $d\equiv 0$. This equation
\[
\partial_t h = D_h \partial_{xx} h + g_h(h), \quad g_h(h)=p_h \frac{r_0 h^2}{ r_1 h^2 +r_3} -\eta_h h,
\]
is in fact a gradient flow related to the free energy $\int D_h \frac{1}{2}h_x^2+ G_h(h)$. 
Here $G_h'=-g$ is the potential with critical points $0,h_\pm$. 
The direction of motion of the front changes sign at the Maxwell point, when $G(h_+)=G(0)$. For the parameter values chosen in our simulations, fronts propagate slowly toward the head region. Fronts between tail and trunk regions are 
completely equivalent. Fronts between tail and head regions do not propagate if they exist. Depending on parameter values, such fronts may not exist but front-like initial conditions rather  split into two fronts $(h_*,0) \leftrightarrow (0,0) \leftrightarrow (0,h_*)$, where the newly emerged state $(0,0)$ expands. 
All of those features can be captured in a scalar equation. 
Somewhat formally, on the level of the kinetics, one can envision straightening out the line segments in the $h-d$-plane $(h_*,0) \to (0,0) \to (0,d_*)$ into a one-dimensional line segment $-1 \to 0 \to 1$; see Figure \ref{f:threetotwo}. 
One then arrives at the scalar equation
\begin{equation}\label{e:order}
 c_t=D_c c_{xx}+c(1-c^2)(c^2-\al^2),
\end{equation}
with $0<\al<1$. Equation (\ref{e:order}) possesses stable 
equilibria $\{-1,0,1\}$ corresponding to $(1,0)$, $(0,0)$, and $(0,1)$, 
respectively, and possesses reflection symmetry between $c\to -c$, mimicking  the reflection symmetry between $d$ and $h$ in our simplified model. Here, the critical ``Mawell'' point is $\al_*=1/\sqrt{3}$, such that $(0,0)$ invades $(1,0)$ and $(0,1)$ for $\al<\al_*$ and $(0,0)$ is invaded otherwise. Similarly to the systems case \eqref{hdw},
 fronts between $-1$ and $1$ alias head and tail exist in \eqref{e:order} precisely when splitting of the front is not expected, that is, when $-1$ invades $0$. This can be readily seen from the phase portrait of the steady-state equation. 

The dynamics for \emph{wnt}-related signal concentrations need to be modified accordingly, with production and degradation now occurring for $c$ positive and negative, respectively, for instance through a term 
\begin{equation}\label{e:wn}
w_t=D_w w_{xx} + p_w \left(c \chi^\eps_{>0} (c)(1-w)+c \chi^\eps_{<0}(c)w\right).
\end{equation}
Gradient sensing at the body edge is lumped together to yield a production term in the equations for the boundary concentrations $c_\pm$ of the form 
\begin{eqnarray}
\frac{d}{dt}c_\pm       &=&-\frac{1}{\gamma}D_c    \partial_\nu c_\pm  +  +c_\pm(1-c_\pm^2)(c_\pm^2-\al^2)  +  \Psi_c^\pm,\label{e:orderbc} \\
\frac{d}{dt}w_\pm       &=& -\frac{1}{\gamma}D_w    \partial_\nu w_\pm    +p_w\left(c_\pm \chi^\eps_{>0} (c_\pm)(1-w_\pm)+
c_\pm \chi^\eps_{<0}(c_\pm)w_\pm\right),\label{e:wnbc}
\end{eqnarray}
where 
\[
\Psi_c^\pm=\tau\chi_{>\theta}^\eps (-\partial_\nu w|_{\pm L})(1-c_\pm)+\tau \chi_{<-\theta}^\eps (\partial_\nu w|_{\pm L})(-1-c_\pm).
\]
The set of equations \eqref{e:order}--\eqref{e:wn} with dynamic boundary conditions \eqref{e:orderbc}--\eqref{e:wnbc} form the minimal reduced system that is able to mimic robust regeneration under cutting, grafting, and growth, which we shall demonstrate in the next section.

\subsection{Homeostasis, cutting, and grafting in the reduced model}\label{s:4.3}
\begin{figure}[h!]
\includegraphics[width=0.27\textwidth]{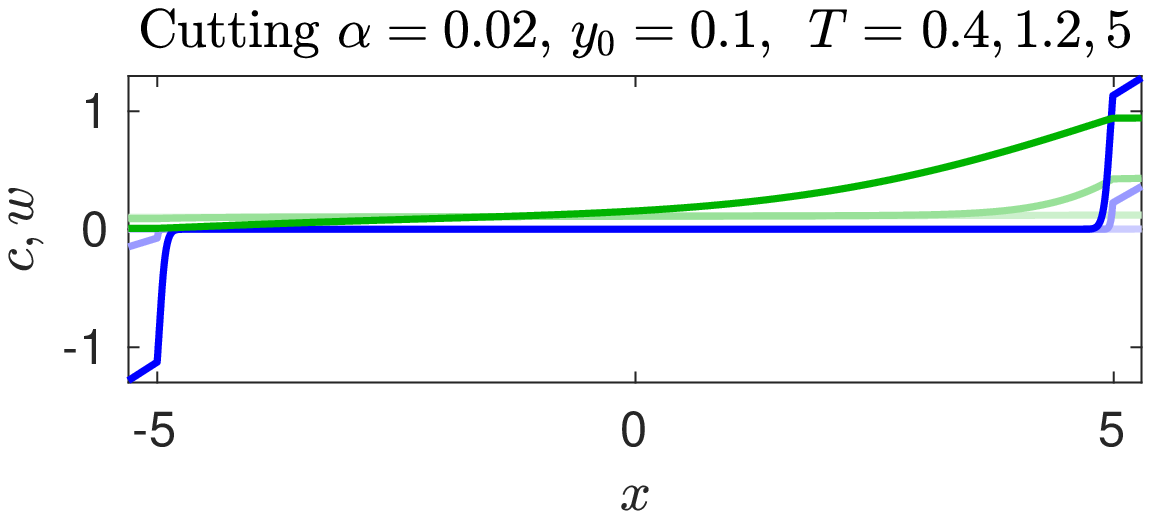}\hfill
\includegraphics[width=0.27\textwidth]{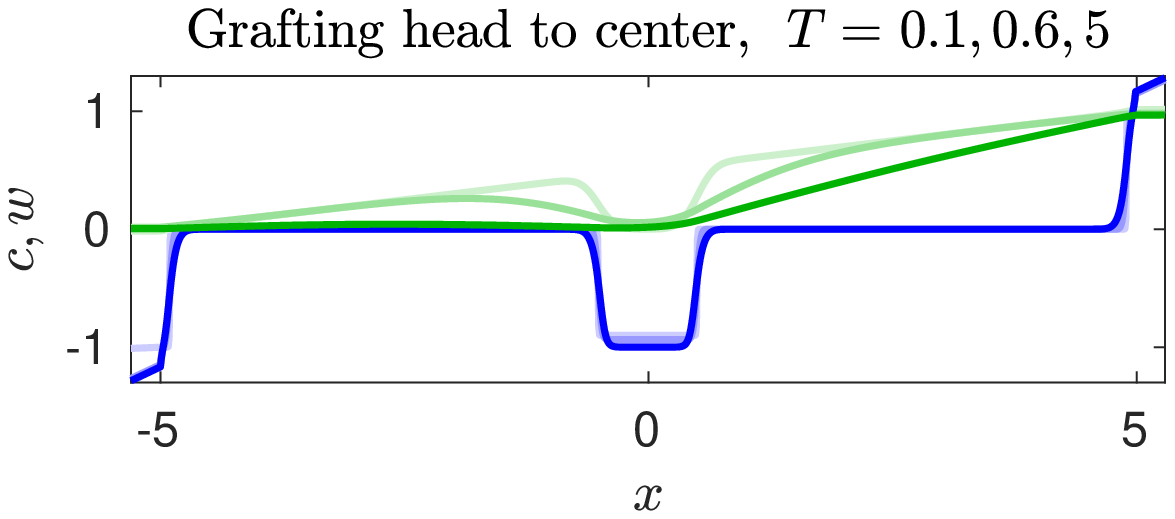}\hfill 
\includegraphics[width=0.27\textwidth]{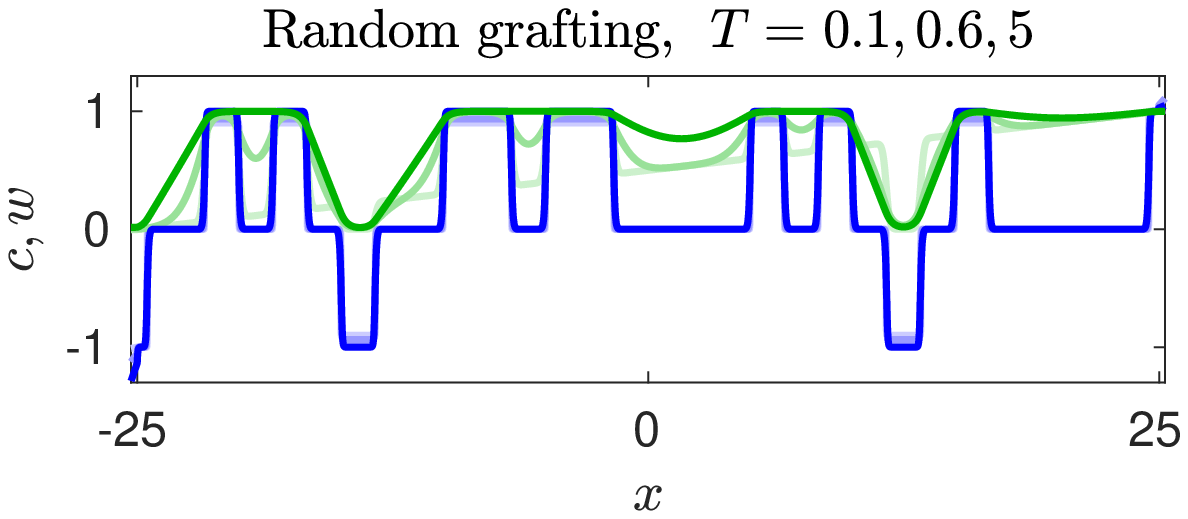}\hfill 
\includegraphics[width=0.129\textwidth]{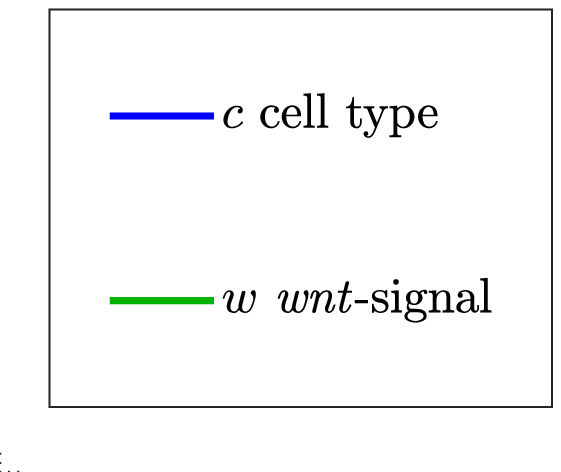}
 \caption{Simulations of the reduced model \eqref{e:order}---\eqref{e:wnbc} with initial conditions corresponding to a cutting experiment (left), a grafting experiment where a head is grafted into the center portion (middle), and a random grafting of head and tail pieces (right); see text for details on initial conditions. }\label{f:red}
\end{figure}

We simulate system  \eqref{e:order}---\eqref{e:wnbc} with parameter values
\[
D_u     = 0,001, \              
D_w     = 1,\  
\al       = 0.577,   \            
p_w     = 10,       \         
\eps     = 0.002,\              
\tau     = 0.5,\                
\theta     = 3,\              
\gamma      = 0.3,\                 
\]
on a domain of size $2L=10$. Figure \ref{f:red} illustrates the prototypical experiments and regeneration with these parameter values. Initial conditions for 
cutting are 
\begin{equation}\label{e:iccut}
w=\frac{\alpha (x+L)}{2L}+y_0 \   , \    c\equiv 0.
\end{equation}
Initial conditions for grafting are the homeostatic equilibrium with 
$w=0$ and $c=\pm 1$ on segments of length $1$. 

Note that we chose $\al$ near the Maxwell point, which prevents changes of size in grafting experiments. Different choices lead to expanding or retracting head- or tail regions, as shown in the full model in Figure \ref{f:homeo2}. 

\begin{figure}[h!]
\includegraphics[width=0.27\textwidth]{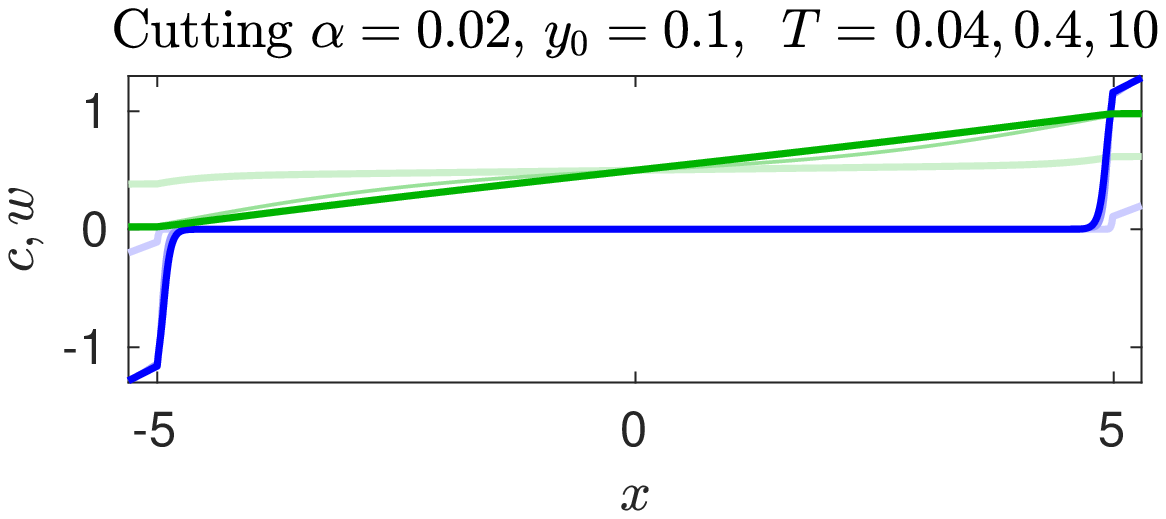}\hfill
\includegraphics[width=0.27\textwidth]{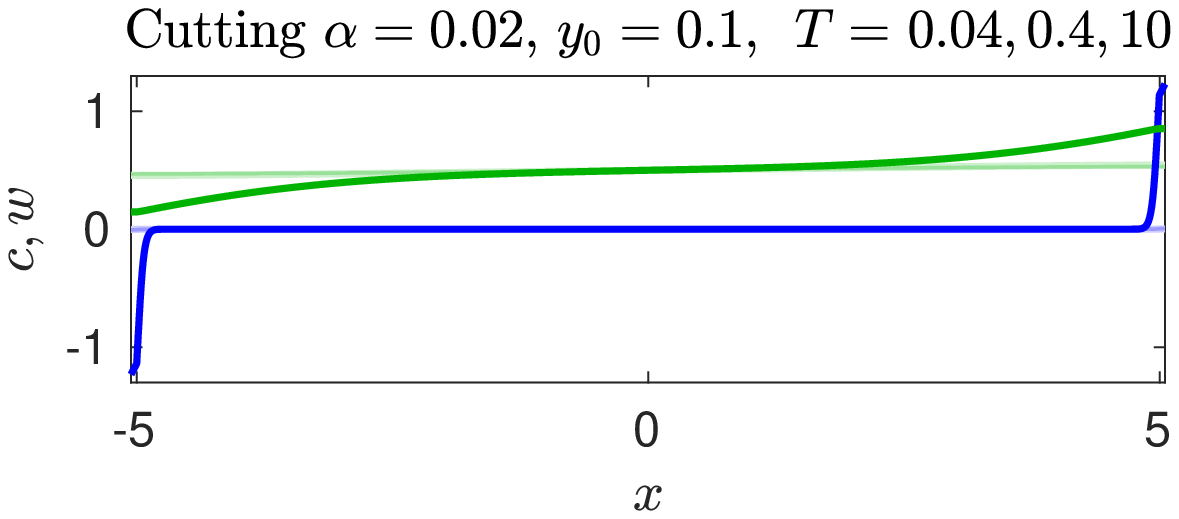}\hfill 
\includegraphics[width=0.27\textwidth]{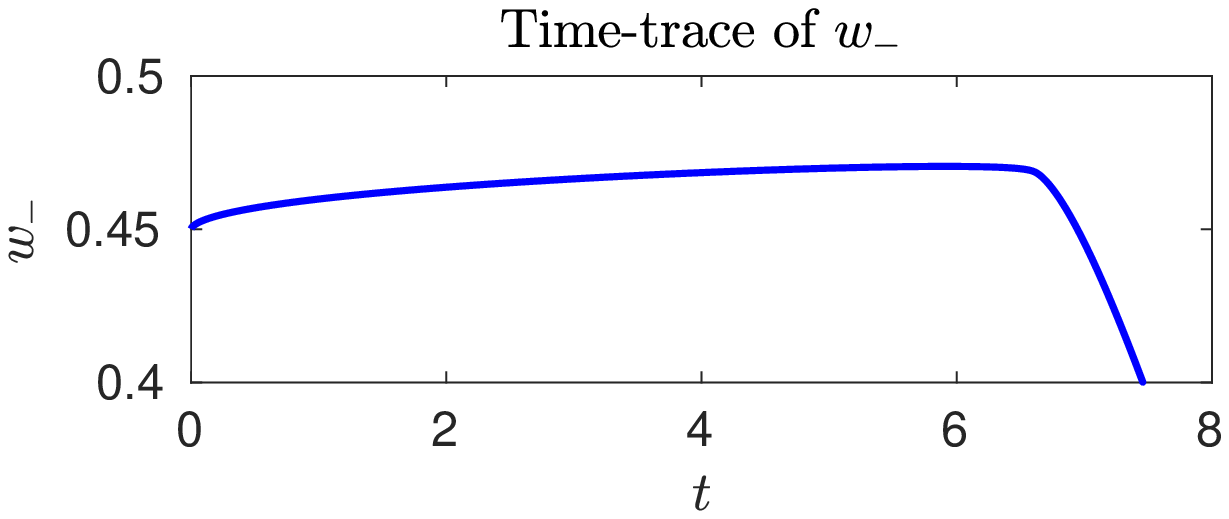}\hfill 
\includegraphics[width=0.129\textwidth]{reduced_legend.eps}
 \caption{Simulations of the reduced model \eqref{e:order}---\eqref{e:wnbc} with initial conditions corresponding to cutting experiments, with parameter choices causing difficult recovery or recovery failure. Larger $\eps=0.012$ leads to slow and non-monotone recovery (left; associated time series of $w_-$ on the right): the \emph{wnt}-signaling gradient first decreases before head and tail cell populations are established. The same phenomenon occurs when reducing the mass in the boundary compartment $ \gamma=0.055$ while keeping $\eps=0.002$ (center). Larger values of $
\eps$ and smaller values of $\gamma$ typically prevent recovery. }\label{f:red2}
\end{figure}

Increasing $\eps$ or decreasing the mass in the body edge compartments $\gamma$ significantly will prevent recovery. Near critical values, we observe non-monotone behavior of the gradient and the boundary values $w_\pm$; see Figure \ref{f:red2}. We will point to some explanation in the next section when we discuss an even further reduced, scalar model. 
As in the full model, we see a dichotomy for head grafts, between merging 
of the two heads for grafts near the head of the host, and persistence 
of one head only for grafts sufficiently far away from the head of the host. 
This we relate to the dichotomy between vanishing of the graft and outgrowth of a new head in experiments. 

\subsection{Analysis and comparison with Robin boundary conditions.}\label{s:redrob} Throughout this section, we restrict to sensitivity thresholds $\theta=0$. 
System \eqref{e:order}--\eqref{e:wn} allows for a trivial solution 
$c\equiv 0$, $w\equiv 1/2$. In fact, cutting experiments with small fragments can be thought of as starting with initial conditions close to this trivial solution. Regeneration and regeneration failure from small fragments can therefore be well predicted from a stability analysis of this trivial solution: stability would prevent regeneration 
from small fragments, while instability indicates the initial phase of the pattern-forming process. 

Linearizing at the trivial solution, we find the system 
\begin{eqnarray}
c_t &=& D_C c_{xx}-\al^2 c, \ |x|<L\\
w_t &=& D_w w_{xx}+\frac{1}{2} c, \ |x|<L\\
\frac{\rmd}{\rmd t} c_\pm &=& -\frac{1}{\gamma}D_c \partial_\nu c|_{\pm L} - \al^2  c_\pm + \frac{\tau}{\eps} \partial_\nu w|_{\pm L},\\
\frac{\rmd}{\rmd t} w_\pm &=& -\frac{1}{\gamma}D_w \partial_\nu w|_{\pm L} +\frac{1}{2}c_\pm.
\end{eqnarray}
After Laplace transform, the associated eigenvalue problem can be converted into a transcendental equation which however is difficult to analyze theoretically. Numerically, one readily finds instabilities in the relevant parameter regimes. We illustrate here the basic instability mechanism with some formal simplifications. 

Consider the left boundary, such that the eigenfunctions are of the form 
$c=c_0 \rme^{-\rho x}$, $w=w_0 \rme^{-\tilde{\rho} x}+ w_1 \rme^{-\rho x}$, where $\rho$ and $\tilde{\rho}$ are determined by the spectral parameter $\lambda$, $D_c \rho^2=\al^2 +\lambda$, $D_w\tilde{\rho}^2=\lambda$, and $w_1$ depends on $c_0$. Substituting these expressions into the equations on the boundary yields a generalized eigenvalue problem. 
\begin{eqnarray}
\lambda c_0 &=& -\frac{1}{\gamma}D_c(-\rho c_0)- \al^2  c_0 + \frac{\tau}{\eps} (-\tilde{\rho}w_0-\rho w_1),\\
\lambda w_0 &=& -\frac{1}{\gamma}D_w (-\tilde{\rho}w_0-\rho w_1) +\frac{1}{2}c_0.
\end{eqnarray}
The key ingredient here are the off-diagonal terms: $\frac{\tau}{\eps}\tilde{\rho} w_0$ from $ \frac{\tau}{\eps} \partial_\nu w$ in the equation for $c_0$, and $\frac{1}{2} c_0$ in the equation for $w_0$. For positive $\lambda$, these terms are both positive hence causing a negative determinant and hence instability. 

These off-diagonal terms can also be understood more directly but less quantitatively  as encoding a positive feedback mechanism. 
A small increase of $w_-$ will cause a positive normal derivative of $w$ and hence an increase in $c_-$ at the boundary from the equation for $c_-$. The increase in $c_-$ translates into a further increase of $w_-$ through the term $\frac{1}{2}c_-$ in the equation for $w_-$, thus providing the positive feedback mechanism responsible for instability.

It is here that we can see how dynamic boundary conditions are a key ingredient. Relaxing these dynamic boundary conditions by letting reaction rates at the boundary tend to infinity, we formally obtain the mixed Robin boundary conditions 
\[
0=-\al^2 c_\pm + \frac{\tau}{\eps} \partial_\nu w|_{\pm L},\qquad 0 = c_\pm. 
\]
One readily sees that, for well-posedness, the condition from the equation for $c_-$ needs to be associated with the equation for $w$ since it contains the $w$ normal derivative, while the condition from the second equation for $c_-$ simply gives a Dirichlet condition for the $c$-equation. As a consequence, the $c$-equation decouples as a simple diffusion equation with decay $-\al^2 c_-$ and Dirichlet boundary conditions. Setting $c=0$ then gives a diffusion equation for $w$ with homogeneous Neumann boundary conditions, which implies stability. 

More directly, this calculation points to the failure of a much simpler model. Indeed, we could model an instantaneous adaptation of tail and head cell concentrations, alias the order parameter $c$, to the sign of the signal gradient through $c=  \mathrm{sign}(\partial_\nu w)$, with a possibly smoothed out version of the sign-function.  One would then let $w$ follow head and tail cell concentrations encoded in $c$, for instance through $w=c$. This approach is bound to fail, since the boundary condition $c=  \mathrm{sign}(\partial_\nu w)$ will \emph{not} enforce $c$ to increase from 0 to 1, say, in a cutting experiment where $\partial_\nu w>0$. Instead, the boundary condition can be satisfied by an, in comparison, much smaller adjustment of the $w$-levels near the boundary which achieves $\partial_\nu w=0$, lowering for instance the concentration of $w$ only slightly in a region close to the boundary. The value  $\partial_\nu w=0$ then is compatible with $c=0$ in the boundary condition (with the convention $\mathrm{sign}(0)=0$). We tested such boundary conditions numerically and observed a subsequent decay of the $w$-gradient and convergence to $c=0$, that is, failure of regeneration. Relating to the previous discussion, the boundary condition $c= \mathrm{sign}(\partial_\nu w)$ fails to enforce regeneration since it is not explicitly associated with a forced change in levels of $c$, but acts rather as a nonlinear flux for the $w$-equation. Dynamic boundary conditions provide precisely this association and therefore guarantee regeneration. 

We numerically confirmed these observations on the necessity of dynamic boundary conditions. We observed stability both in a discretization with mixed boundary conditions and when reducing mass fractions to the scale of the grid size.

\subsection{Recovery of the long-range signal gradient as organizing 
feature}\label{s:4.4}
We can achieve a further simplification to a single scalar equation if we focus on experiments with only head and tail at the respective extremity, excluding in particular grafting experiments, and tracking only \textit{wnt}-related concentrations. We assume that the order parameter $c$ mostly vanishes in the domain, that is, head- and tail-cells are confined to the boundary regions. In this case, the kinetics for $w$ vanish inside the bulk $|x|<L$, and we are left with a simple diffusion equation for $w$. Note that his explicitly excludes grafting experiments and more generally states with multiple head and tail regions. 

With this assumption it is also sufficient to track the boundary data $c_\pm$ for the order parameter. Assuming now that $c_\pm$ adjusts rapidly according to \eqref{e:orderbc}, we can  set $c_\pm\sim \mathrm{sign}\,\partial_\nu w$. This all gives us the simple scalar equation 
\begin{equation}
w_t=D_w w_{xx},\ |x|<L,\qquad \frac{d}{dt}w_\pm   =
-\frac{1}{\gamma}D_w    \partial_\nu w|_{\pm L}  +   \Psi_w^\pm \   ,
\label{e:orderscal}
\end{equation}
with 
\[
\Psi_w^\pm=\tau\left[\chi^\eps_{<-\theta}(\partial_\nu w)(-w)
+\chi^\eps_{>\theta}(\partial_\nu w)(1-w)\right].
\]

\begin{figure}[h!]
\includegraphics[width=0.27\textwidth]{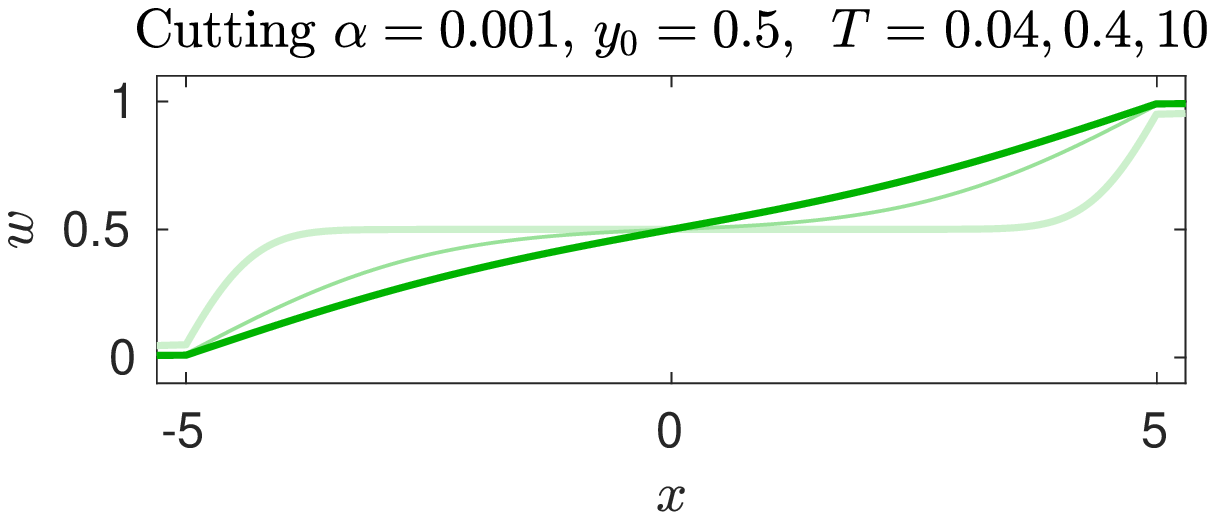}\hfill
\includegraphics[width=0.27\textwidth]{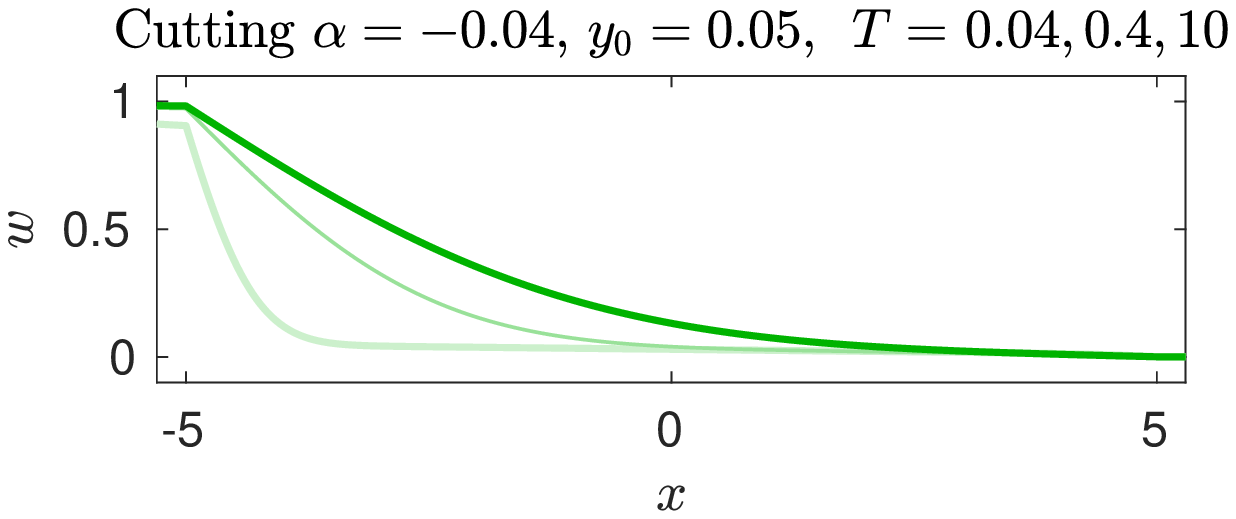}\hfill 
\includegraphics[width=0.27\textwidth]{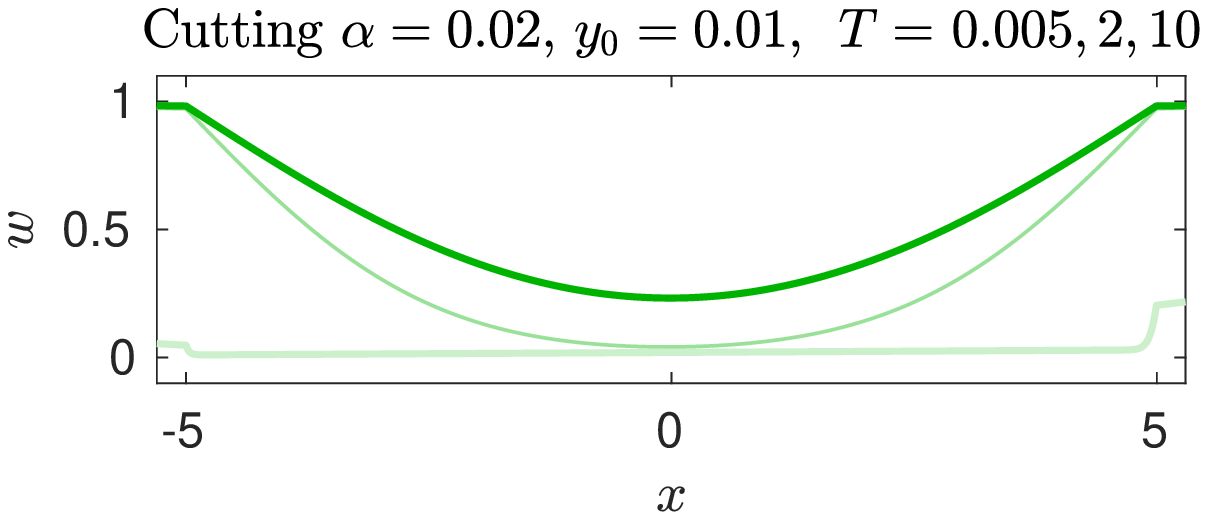}\\
\includegraphics[width=0.27\textwidth]{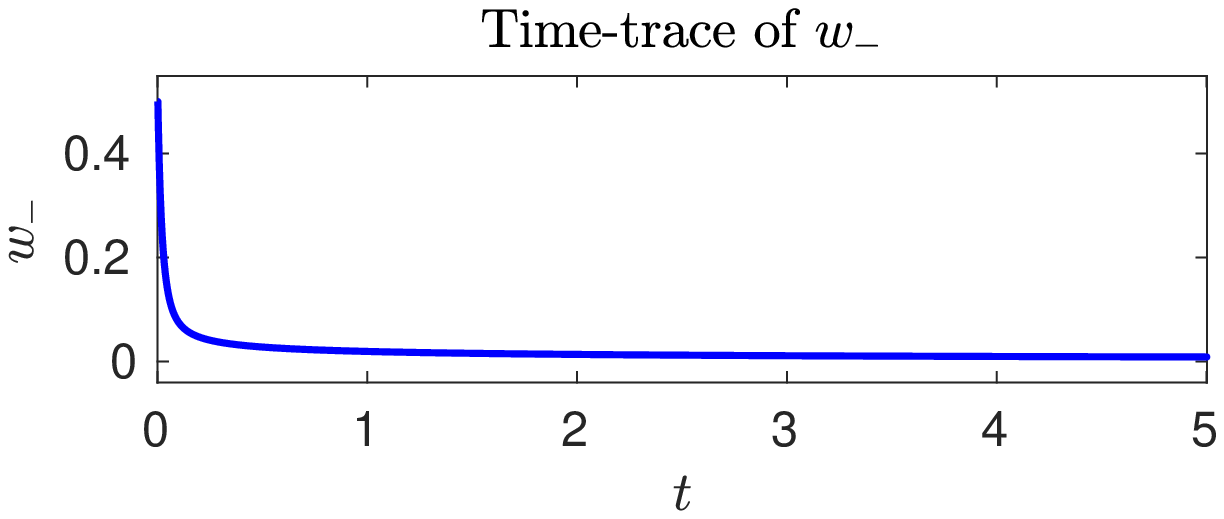}\hfill
\includegraphics[width=0.27\textwidth]{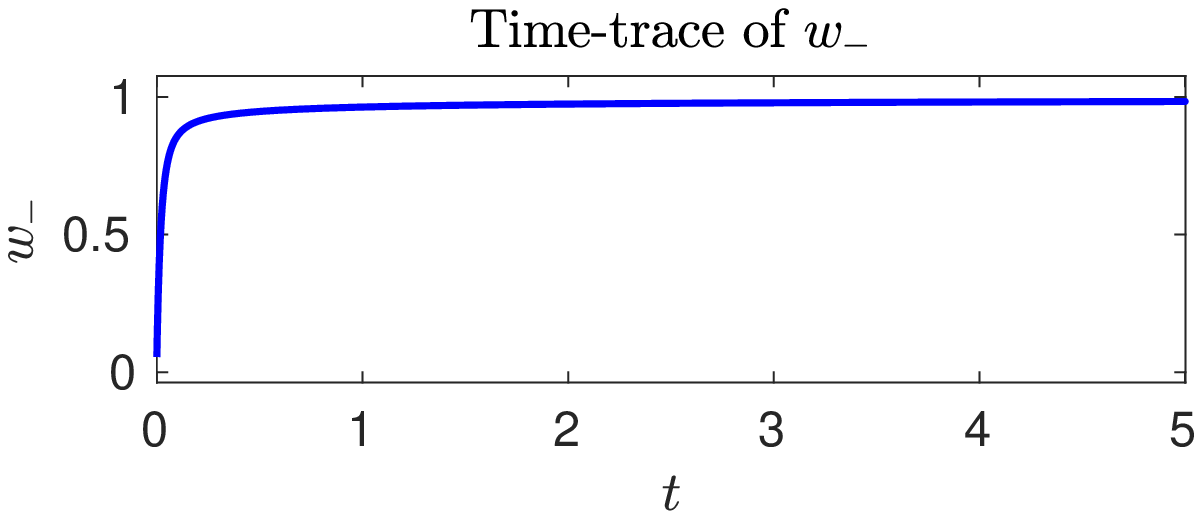}\hfill 
\includegraphics[width=0.27\textwidth]{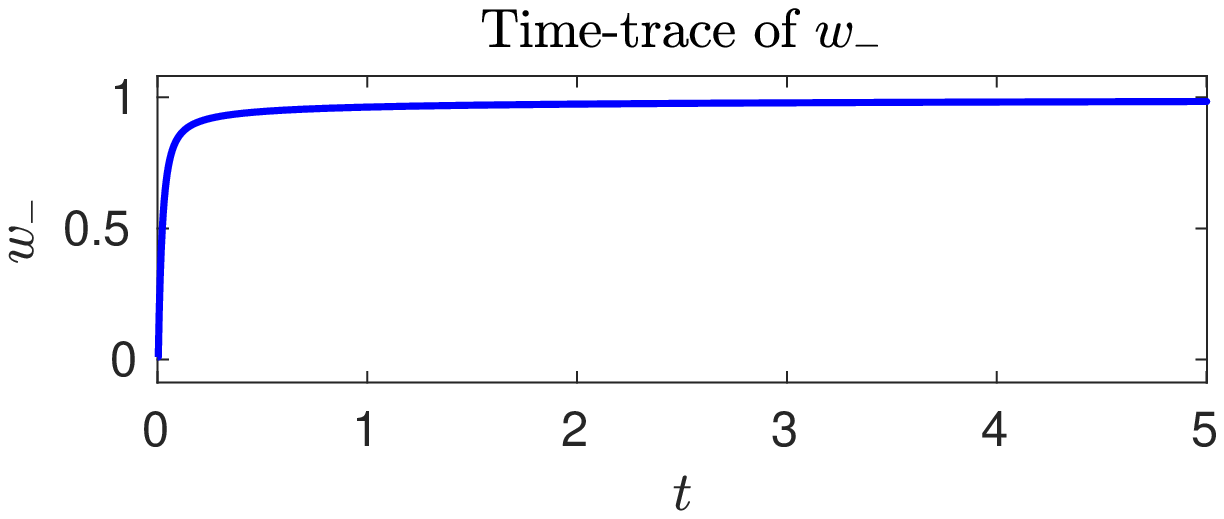}
 \caption{Simulations of the scalar model \eqref{e:orderscal} with initial 
conditions corresponding to cutting experiments, that is, linear profiles in 
$w$ with small slopes. The left panel shows robust recovery for a very small fraction, $0.1\%$, from the center region. The center panel shows recovery for a somewhat larger fraction, $4\%$, from the head region. Here the cut is reflected to demonstrate that recovery is independent of the orientation of the $x$-interval. The right panel shows failure of recovery when a slightly smaller fragment is cut from the head region, in agreement with experimental observations that recovery from fragments near head and tail is less robust; see text for details.   Note the slightly different time instances shown in the right panel, illustrating the very rapid failure to preserve the sign of $\partial_\nu w$ near the boundary. 
The second row shows corresponding time 
series of $w_-$\ . }\label{f:scal1}
\end{figure}

\begin{figure}[h!]
\includegraphics[width=0.27\textwidth]{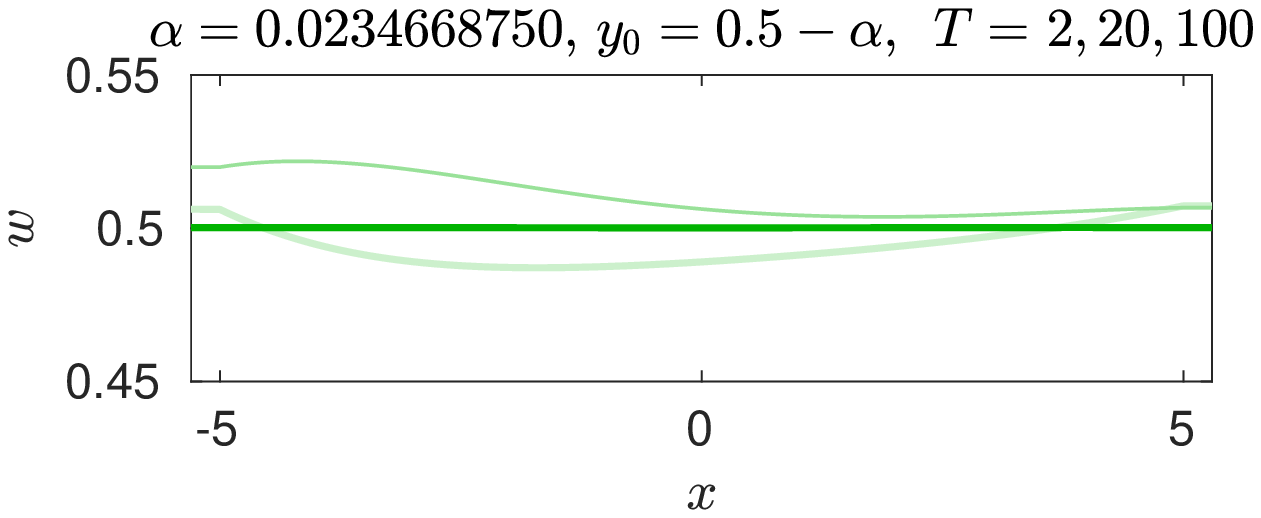}\hfill 
\includegraphics[width=0.27\textwidth]{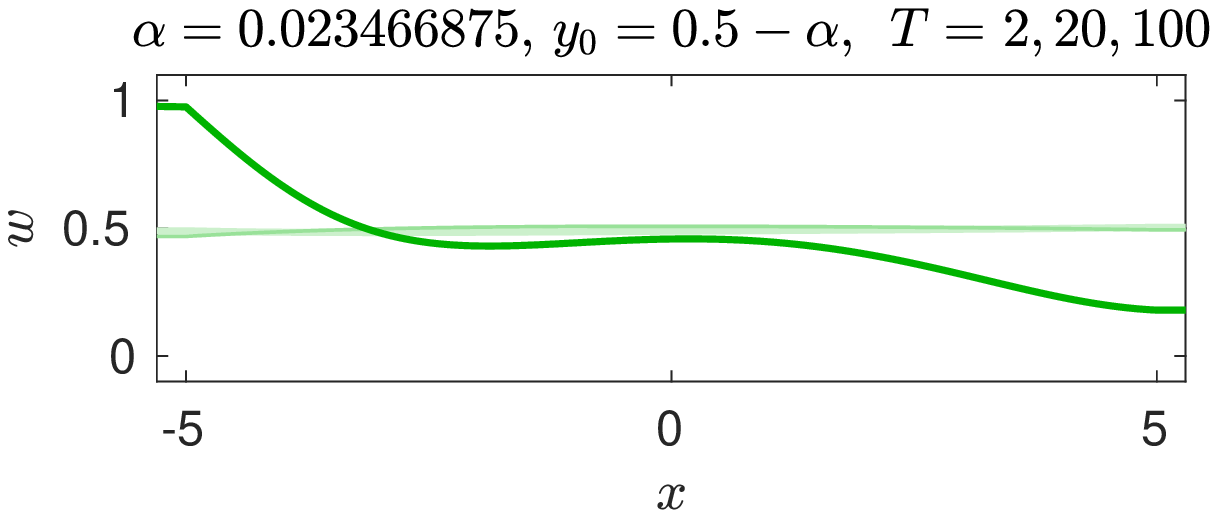}\hfill
\includegraphics[width=0.27\textwidth]{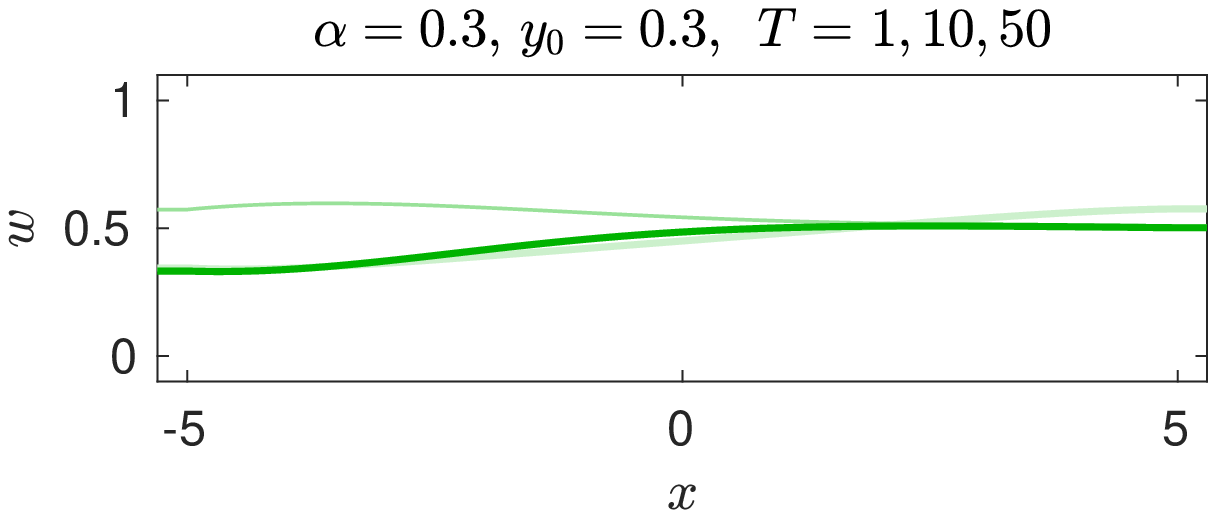}\\
\includegraphics[width=0.27\textwidth]{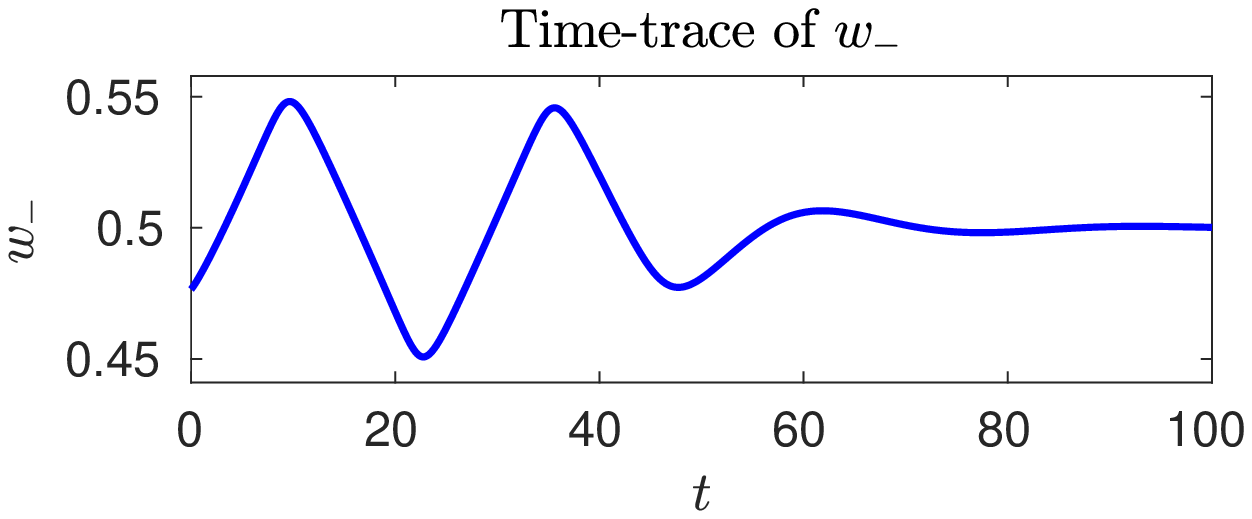}\hfill 
\includegraphics[width=0.27\textwidth]{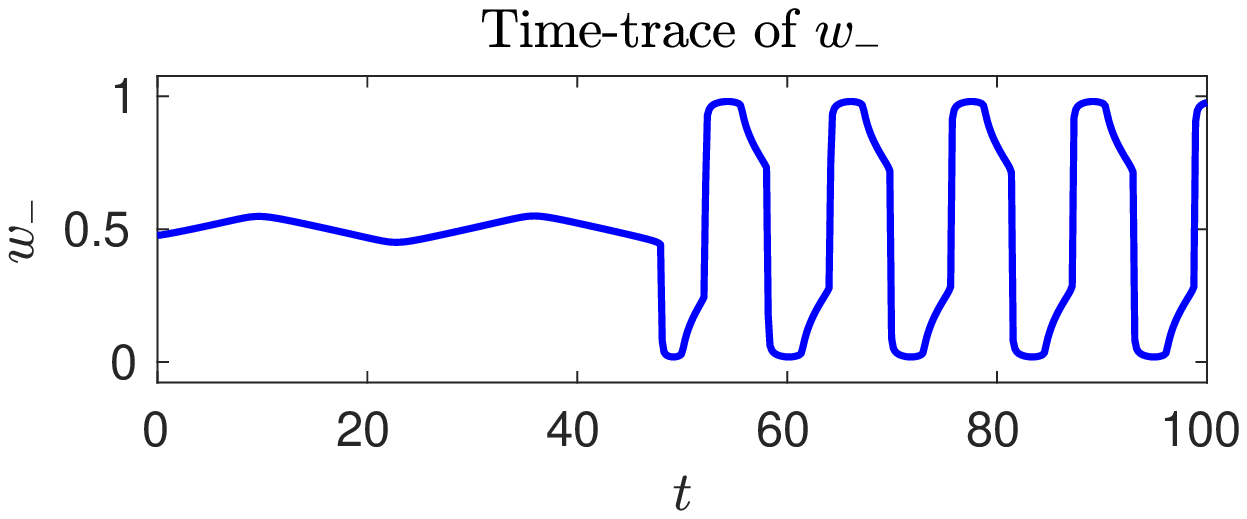}\hfill
\includegraphics[width=0.27\textwidth]{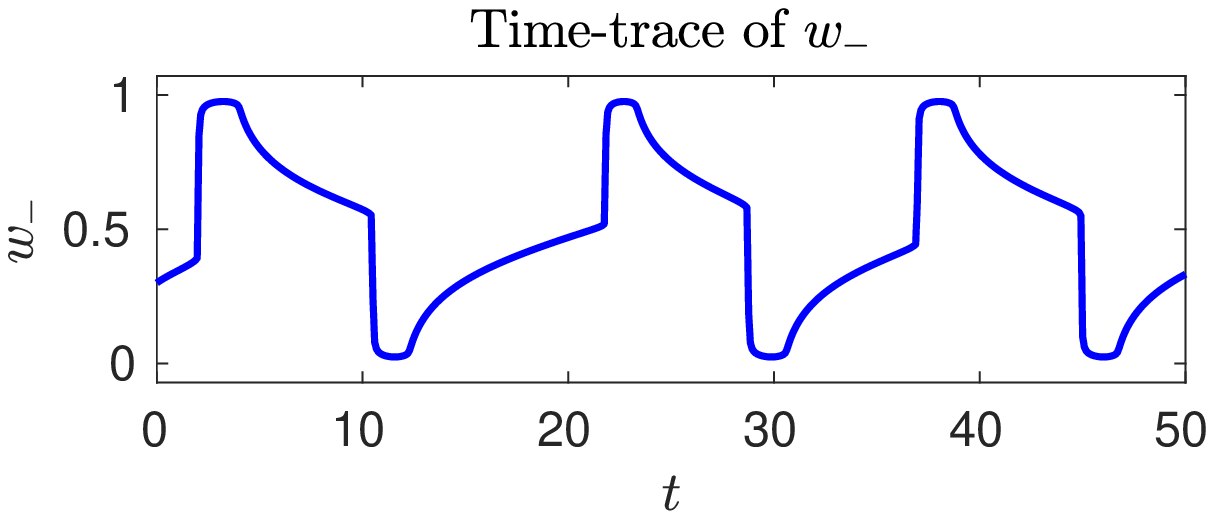}
 \caption{Simulations of the scalar model \eqref{e:orderscal} with initial 
conditions corresponding to a cutting experiments and moderate value of 
sensitivity $\eps^{-1}=15$. An unstable oscillation separates initial 
conditions w.r.t.  $\alpha$ that decay to $w=0.5$ (left) from initial 
conditions with sustained large-amplitude oscillations (center). Note the 
different scales for $w$ here. Oscillations disappear in a saddle-node of 
periodic orbits for smaller values of $\eps$. For yet smaller sensitivity
$\eps\sim 12.5$, the oscillations seem to disappear in a heteroclinic bifurcation (right).
 Corresponding time series of $w_-$ in the second row. See text for predictions of Hopf bifurcations and implications for model corroborations.  Oscillations are 
illustrated in the supplementary materials \textsc{scalar\_oscillations.mp4}.} 
\label{f:scal2}
\end{figure}

\paragraph{Cutting --- numerical experiments.}
We first illustrate regeneration of the \emph{wnt}-signaling gradient after cutting, and therefore use parameters 
\[
D_w     = 1,\ p_w     = 10,\ \eps     = 0.002,\ \tau     = 50,\ \theta= 3,\ \gamma      =0.3,\ 
L       = 10.
\]
Figure \ref{f:scal1} shows recovery from small fragments cut from center- or head regions. Recovery is slightly less robust for fragments cut near head or tail. 
For increasing values of $\eps$ which amounts to decreasing the sensitivity of the gradient sensing at the boundary (or, also, for increasing values off the mass fraction $\gamma$), we see a transition to a system state which fails to recover; see Figure \ref{f:scal2}. 
One first sees oscillatory decay toward the trivial state $w\equiv 0.5$ before sustained oscillations emerge in a weakly subcritical Hopf bifurcation. The unstable periodic orbit separates initial conditions that lead to the trivial, trunk-only state $w\equiv 0.5$ from initial conditions that converge to large-scale oscillations. The large oscillations eventually appear to terminate in a heteroclinic bifurcation for larger values of $\eps$. We saw qualitatively similar transitions when decreasing the mass fraction $\gamma$ instead of increasing $\eps$.

\paragraph{Analysis --- equilibria and stability.}
We assume that $\tau\gg 1$ and neglect the flux term $-\frac{1}{\gamma}D_w    \partial_\nu w$. We also assume that $\theta=0$ such that the characteristic function takes value $1/2$ at $w=1/2$. At the boundary, equilibria then satisfy either,
\[
\bullet\ w=1, \ \partial_\nu w>0;\qquad
\bullet\  w=0, \partial_\nu w<0;\qquad
\bullet\ w=1, \partial_\nu w =0.
\]
For moderately sized domains, $L\ll \eps$, this yields three equilibria, $w\equiv 1/2$,  $w=(x+L)/(2L)$, and $w=(-x+L)/(2L)$, corresponding to only trunk, head-tail, and tail-head solutions. 
Linearizing at these equilibria, we find a diffusion equation with dynamic boundary conditions
\begin{equation}
w_t=D_w w_{xx},\ |x|<L,\qquad \frac{d}{dt}w_\pm   =-\frac{1}{\gamma}D_w    
\partial_\nu w|_{\pm L}  -\tau w + \mu \frac{\tau}{\eps} \partial_\nu w \   , 
\label{e:orderscal2}
\end{equation}
where $\mu=1$ for the trunk solution $w\equiv 0$ and $\mu=0$ for the head-tail and tail-head solutions. 

One readily finds that head-tail and tail-head solutions are always stable. 
To analyze stability of the trunk-only solution, $mu=1$, we consider the semi-unbounded domain $x>0$, set $w=\rme^{\lambda t + \rho x}$, with $\lambda=\rho^2$, to find the characteristic equation 
\[
\rho^2 + \left(\frac{\tau}{\eps}-\frac{1}{\gamma}\right)\rho +1=0.
\]
For $\rho$ to correspond to an eigenvalue, we need $\Re\rho<0$, which implies $\frac{\tau}{\eps}>\frac{1}{\gamma}$. We find
\begin{itemize}
\item \emph{real instability:} $\frac{\tau}{\eps}>\frac{1}{\gamma}+2$ gives 2 real roots $\rho_\pm<0$ and two associated real unstable eigenvalues $\lambda_\pm>0$;
\item \emph{complex instability:} $\frac{1}{\gamma}+2>\frac{\tau}{\eps}>\frac{1}{\gamma}+\sqrt{2}$ gives 2 complex conjugate roots with $\Re\rho_\pm<0$ and two associated complex conjugate unstable eigenvalues $\Re\lambda_\pm>0$;
\item \emph{Hopf bifurcation:} $\frac{\tau}{\eps}=\frac{1}{\gamma}+\sqrt{2}$ gives 2 complex conjugate roots with $\Re\rho_\pm<0$ and two purely imaginary eigenvalues $\Re\lambda_\pm=0$;
\item \emph{stability:}  $\frac{\tau}{\eps}<\frac{1}{\gamma}+\sqrt{2}$ gives eigenvalues $\lambda$ with negative real part.
\end{itemize}
We found that the Hopf bifurcation is subcritical, the unstable periodic solution stabilizes in a saddle-node bifurcation of periodic orbits, grows in amplitude and eventually disappears in a heteroclinic bifurcation for sufficiently large values of $\gamma$.

The dynamics are quite different if the boundary conditions are relaxed to Robin boundary conditions, for instance 
\[
 \tau\left(\chi^\varepsilon_{<-\theta}(\partial_\nu w)(-w)
+\chi^\varepsilon_{>\theta}(\partial_\nu w)(1-w)\right)=0,
 \]
 at $x=\pm L$. Neglecting the flux term $-\frac{1}{\gamma}D_w    \partial_\nu w$, that is, with sufficiently large mass fractions $\gamma$, equilibria for dynamic and Robin boundary conditions coincide. Stability of equilibria is however quite different: the linearization with Robin boundary conditions is a 
Sturm-Liouville eigenvalue problem
\[
\lambda w=w_{xx},\ x\in (-L,L),\qquad \frac{\mu}{\varepsilon}\partial_\nu w-w=0,\ |x|=L, 
\]
with real eigenvalues $\lambda_0>0>\lambda_1>\ldots$, $\lambda_0\sim \eps^2/\mu^2$ for $L\gg 1$. 
The dynamics 
generally do not allow for oscillations as observed in the case of periodic boundary conditions. Nevertheless, the trivial trunk-only solution is unstable in this approximation and we see robust recovery of small $w$-gradients. One can attribute the appearance of oscillations to an effective distributed delay in the otherwise scalar equation for the boundary dynamics of $w_-$ caused by the coupling to the diffusive field $w(t,x)$ which acts as a buffer that stores a blurred history of boundary data. 
 
 The transitions discussed here occur when gradient dependence at the boundary is not sufficiently sensitive, that is, $\eps$ is not sufficiently small or $
\tau$ is too small, or when the mass fraction in the boundary region $\gamma$ is too small. This appears to be a prediction quite specific to this model, occurring to some extent  also  in the full system and the system with order parameter. We are not aware of experimental observations of oscillations in cases where recovery and regeneration are severely impeded. But any such observation would clearly corroborate our basic modeling assumptions. 

\section{Discussion}

We presented a mathematical  model for robust regeneration of planarians. Our model is able to reproduce most cutting and grafting experiments. As opposed to modeling efforts based on Turing-type mechanisms, our model preserves polarity after cutting and yields robust results over organism scales differing by factors of 100. 

Central to our model are two observations from experiments: 
\begin{itemize}
\item[(i)] sharply increased activity including stem cell proliferation near wound healing sites;
\item[(ii)] global gradients of chemical signals, related to the \emph{wnt}-signaling pathway.
\end{itemize}
We translate the first observation into dynamic boundary condition, modeling changed reaction kinetics in a boundary compartment. The second observation has often been discussed in connection with the regulation of tissue size, expanding on the idea of the French-flag model. The role of this global signal gradient is different in our model: we postulate that the gradient, rather than absolute levels of the signal are sensed by stem cell populations and, at wound sites, translated into directed differentiation. We suspect that within a rather general modeling context, such a gradient sensing is necessary in order to reproduce robust preservation of polarity in cutting experiments. 

We incorporate these ingredients into a comprehensive model for 6 species, 3 cell type populations and 3 chemical signals.  Through model reduction to an order parameter for cell types and one long-range chemical signal, only, we exhibit how these two ingredients organize the regeneration process. In the reduced model, we can point to regeneration as an instability mechanism for a trivial, 
unpatterened state and identify analytically limits of robust regeneration. The process is fundamentally different from Turing's mechanism, and driven by the boundary compartments.

There are several ways in which the model could be refined. First, one can quite easily introduce a head-tail bias, which is clearly observed in experiments. This could be introduced either through different sensing thresholds $\theta$ at body edges, or through different differentiation and proliferation kinetics. One can thus introduce a bias towards head generation from small fluctuation of 
trivial states. We are not aware of a causal rather than a phenomenological justification of this bias. Second, we do not attempt to model regulation of the 
size of the head and tail regions. Since our model relies essentially on establishing a global signaling gradient, cells can clearly obtain positional information by reading out absolute levels of the \emph{wnt}-related signal. Postulating such a dependence of differentiation or apoptosis on these levels, one can then introduce $w$-dependence in the $\{s,h,d,u_h,u_d\}$-subsystem and thus influence the tristability. Front motion as described in the reduced model in Section \ref{s:4.2} would then depend on $w$ levels, and stationary interfaces would lock into fixed $w$-levels, thus regulating a fixed size of head or tail as a percentage of the full body length. In the order parameter model, this could for instance be accomplished by a $w$ dependence of the quintic kinetics,
\[ 
c_t= c_{xx}+c(1-c^2)(c-\chi^\eps_{>0.85}(w))(c-\chi^\eps_{<-0.85}(w)),
\]
which would regulate the size of head and tail to regions where $w>0.8$ and $w<-0.8$, respectively, about 15\% of body size, each. Pushing this further, such effects could also model the spontaneous formation of head cell clusters when suppressing the \emph{wnt}-signaling pathway.
Lastly, one could address directed motion of stem cells or progenitors. Our 
focus here was on essential features enabling regeneration in a minimal model 
and we therefore ignored (active) transport other than random motion and diffusion. Since experiments appear to point to a significant role of directed motion, a model that takes such effects into account could potentially improve both qualitatively and quantitatively on the results presented here. 
%
%

\end{document}